\documentclass{IEEEtran4PSCC}
\ifCLASSINFOpdf
   \usepackage[pdftex]{graphicx}
\else
   \usepackage[dvips]{graphicx}
\fi
\usepackage[cmex10]{amsmath}
\usepackage{array}
\makeatletter
\let\old@ps@headings\ps@headings
\let\old@ps@IEEEtitlepagestyle\ps@IEEEtitlepagestyle
\def\psccfooter#1{%
    \def\ps@headings{%
        \old@ps@headings%
        \def\@oddfoot{\strut\hfill#1\hfill\strut}%
        \def\@evenfoot{\strut\hfill#1\hfill\strut}%
    }%
    \def\ps@IEEEtitlepagestyle{%
        \old@ps@IEEEtitlepagestyle%
        \def\@oddfoot{\strut\hfill#1\hfill\strut}%
        \def\@evenfoot{\strut\hfill#1\hfill\strut}%
    }%
    \ps@headings%
}
\makeatother

\psccfooter{%
        \parbox{\textwidth}{\hrulefill \\ \small{23rd Power Systems Computation Conference} \hfill \begin{minipage}{0.2\textwidth}\centering \vspace*{4pt} \includegraphics[scale=0.06]{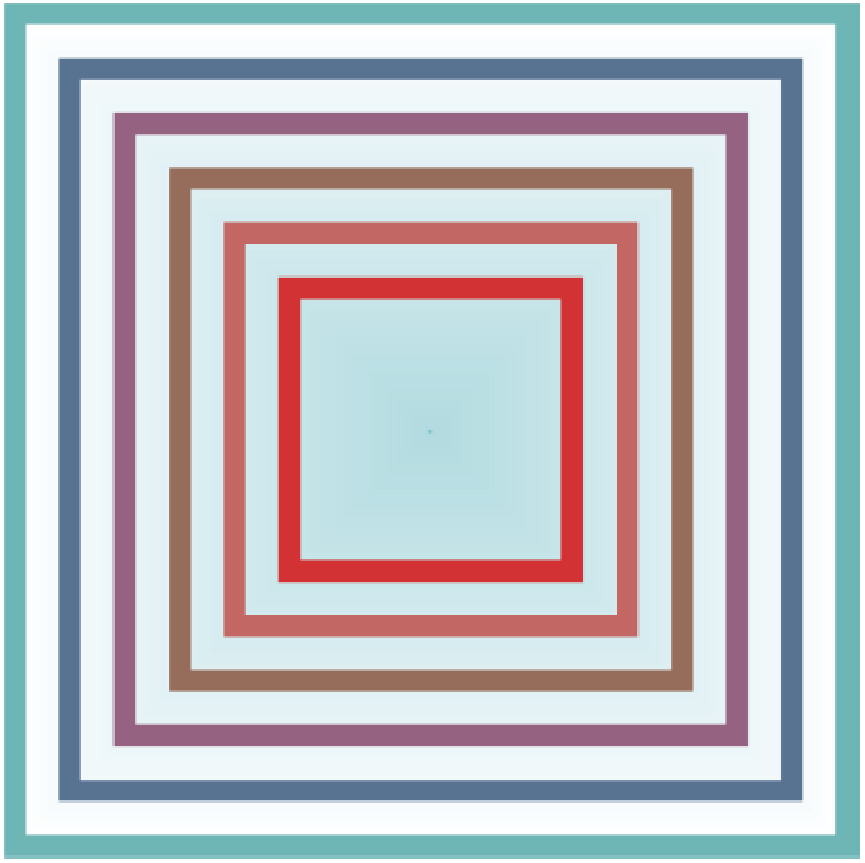}\\\small{PSCC 2024} \end{minipage} \hfill \small{Paris, France --- June 4 -- 7, 2024}}%
}

\usepackage{amssymb}
\usepackage{tikz}
\usetikzlibrary{positioning}
\usepackage{circuitikz}
\usepackage{pstricks}
\usepackage{steinmetz}

\newcommand{\bmat}[1]{\begin{bmatrix} #1 \end{bmatrix}}

\providecommand{\abs}[1]{\lvert#1\rvert}

\newcommand{\pd}[2]{\frac{\partial #1}{\partial #2}}

\begin{document}
%
\title{Toward More Accurate and Robust  \\ Optimal Power Flow for Distribution Systems}

\author{
\IEEEauthorblockN{Dakota Hamilton\\Loraine Navarro\\Dionysios Aliprantis}
\IEEEauthorblockA{Elmore Family School of Electrical and Computer Engineering\\
Purdue University\\
West Lafayette, IN, USA\\
\{hamilt89, navarr50, dionysios\}@purdue.edu}
}


\maketitle

\begin{abstract}
The objective of this paper is to improve the accuracy and robustness of optimal power flow (OPF) formulations for distribution systems modeled down to the low-voltage point of connection of individual buildings.
An approach for addressing the uncertain switching behavior of building loads (e.g., air conditioners, water heaters, or pool pumps) and variable renewable generation (e.g., rooftop solar) in the OPF is proposed.
Rather than using time-averaged forecasts to determine voltage magnitude constraints, we leverage worst-case minimum and maximum forecasts of loads and distributed energy resource generation.
Sensitivities of the power flow equations are used to predict how these deviations in load and renewable generation will impact system voltages, and the voltage constraints in the OPF are dynamically adjusted to mitigate voltage violations due to this uncertainty. 
A methodology for incorporating models of split-phase components and transformer core losses in the OPF formulation is also proposed. 
The proposed approach is validated through numerical case studies on a realistic distribution feeder using GridLAB-D, a distribution system simulation software.
\end{abstract}


\thanksto{\noindent Submitted to the 23rd Power Systems Computation Conference (PSCC 2024).
This material is based on work that was supported by the U.S. Department of Energy (under Award No.\ DE-OE0000921).}

\section{Introduction}
With the growing penetration of distributed energy resources (DER), such as rooftop solar panels, electric vehicles, energy storage, and flexible loads, the landscape of modern power systems is rapidly changing from centralized to distributed.
Furthermore, the proliferation of smart meters and advanced metering infrastructure (AMI) have enabled better data collection, and improved modeling of low-voltage distribution feeders~\cite{Peppanen2014,Morrell2018,Guo2022}.
These factors are pushing operation and control decisions increasingly closer to the grid edge.
Power system optimization problems, such as optimal power flow (OPF), that were once solved at the transmission level may soon become commonplace in advanced distribution management system software.
In this paper, we address some of the unique challenges associated with implementing OPF for unbalanced, low-voltage distribution feeders, while considering modeling fidelity down to the individual building level.

Power distribution networks are typically unbalanced due to unbalanced loads, single- or double-phase laterals, and non-transposed distribution lines.
OPF formulations for unbalanced multi-phase distribution systems have previously been proposed for a variety of applications, including economic dispatch, active distribution grid management, service restoration, and power quality improvement~\cite{Low2014,soares2017active,wang2019coordinating,oikonomou2019deliverable,elkadeem2019optimal,su2014optimal,costa2017multi,brandao2019optimal,liu2020fairness}.  
However, many of these approaches involve aggregating load and DER to medium-voltage nodes. Instead, we focus on modeling distribution feeders down to the low-voltage point of connection of individual buildings, which is important if one wishes to consider the actions of individual end users or prosumers in the analysis. 

Several papers have proposed OPF formulations for low-voltage residential networks~\cite{su2014optimal,costa2017multi,brandao2019optimal,liu2020fairness}. 
In these studies, the load consumption from each individual household is modeled as a constant over a given window of time (e.g., 15 minutes), based on an average forecast. 
However, the switching behavior of residential loads such as thermostatically-controlled loads (e.g., air conditioners or waterheaters), electric vehicles, and pool pumps, are difficult to predict and can cause the actual power consumption of a building to vary drastically from the forecasted average within a given window. 
This switching behavior can result in large fluctuations in voltage magnitudes at low-voltage nodes, which may violate voltage limits.
Additionally, uncertainties in forecasts of available renewable generation, such as rooftop solar, can also contribute to voltage issues in distribution feeders.
Thus, OPF formulations that use constraints (e.g., voltage magnitude limits) based on average load and DER generation forecasts are insufficient to ensure safe system operation. 
Furthermore, we emphasize that it is important to model voltage constraints at the low-voltage point of connection of customers within the network; applying these constraints at the medium-voltage level does not account for the voltage drop across the distribution transformers, which typically dominates the voltage drop of the cables forming the network. 

Various methods for modeling forecast uncertainties in the OPF based on probabilistic constraints have been proposed~\cite{roald2017chance,dall2017chance,lubin2019chance}.
Using chance constraints, we can enforce limits on the probability that voltage magnitudes will violate their bounds. 
However, such approaches rely on knowledge of probability distribution functions for loads and DER generation, which may not be readily available in practice. 
In this paper, we introduce a different approach to account for the switching behavior of residential loads and uncertainty in renewable generation forecasts. 
We propose to set the voltage constraints in the OPF based on worst-case load and DER generation forecasts, which unlike probability distribution functions, can more easily be estimated from limited historical smart meter data.
Specifically, we use sensitivities of the unbalanced, three-phase power flow equations to predict how uncertainty in load and DER generation forecasts will impact system voltages.
We then dynamically reformulate the voltage constraints in the OPF each time the problem is solved to improve its robustness to this uncertainty.

Another challenge in implementing OPF for distribution feeders is  modeling relevant system components in a computationally tractable way.
In this paper, we leverage the three-phase OPF formulation based on the branch flow model (BFM) of the power flow equations~\cite{Low2014}.
At medium-voltage distribution levels, models for including on-load tap-changing transformers, voltage regulators, and devices that provide reactive power compensation (e.g, capacitor banks or smart inverters) in BFM-based OPF formulations have been proposed in the literature~\cite{wu2016exact,nguyen2018exact}. 
When considering low-voltage networks (in the United States) down to the individual building level, it is critical to accurately model split-phase secondaries consisting of center-tapped transformers and triplex lines. 
In particular, the core losses of low-voltage distribution transformers can be significant.
Hence, in this paper, we also extend the unbalanced three-phase OPF formulation based on the BFM to include models of split-phase components and transformer core losses.

In summary, the main contributions of this work are:
\begin{enumerate}
    \item A practicable modeling approach for improving the robustness of the OPF in distribution feeders to uncertainty in load and DER generation forecasts, especially due to the switching behavior of loads. This is based on the sensitivities of the three-phase power flow equations, which are used to adjust voltage magnitude constraints in the OPF to account for forecasting uncertainties.
    \item A computationally tractable formulation of the BFM-based OPF that includes models of split-phase secondaries and center-tapped transformers (including their core losses, which can be significant and should be accounted for judiciously).
\end{enumerate}

The remainder of this paper is organized as follows:
Section~\ref{sec:opf} introduces the formulation of the three-phase OPF based on the BFM.
The mathematical details for incorporating uncertainty-robust constraints in the OPF are provided in Section~\ref{sec:jacobian}, and Section~\ref{sec:split_phase} describes the modeling of split-phase secondaries and center-tapped transformers.
In Section~\ref{sec:results}, we verify the accuracy of component modeling in the proposed approach, as well as demonstrate the reduction of constraint violations due to uncertainty, through numerical case studies of realistic, large-scale distribution feeders using GridLAB-D, a distribution system simulation tool. 
Section~\ref{sec:conclusion} concludes and discusses future research directions.

\section{OPF Problem Formulation}
\label{sec:opf}
We consider a radial, three-phase distribution network, and denote its graph by $\mathcal{G} = \{\mathcal{N},\mathcal{E}\}$.
We denote the set of phases of node $i \in \mathcal{N}$ by $\Phi_i$, and the set of phases of branch $(j,k) \in \mathcal{E}$ by $\Phi_{jk}$.
For example, a two-phase branch with phases $a$ and $c$ would have $\Phi_{jk}=\{a,c\}$.
We leverage the BFM to formulate the three-phase OPF as the following optimization problem:\footnote{For numerical reasons, we convert all quantities in the OPF formulation into per unit.}
\begin{align}
\label{eq:encap_SDP}
    \min~~& C(\mathbf{s}_{i}^G\,,\,\mathbf{V}_{\!i}\,,\,\mathbf{S}_{jk}\,,\,\mathbf{L}_{jk})\,,\\
    \text{over}~~&\mathbf{s}_{i}^G \in \mathbb{C}^{\abs{\Phi_i}}\,,~\mathbf{V}_{\!i}\in\mathbb{H}^{\abs{\Phi_i}\times\abs{\Phi_i}}\,,~~\forall i \in \mathcal{N}\,;~~\abs{V_0} \in \mathbb{R}\,;\nonumber\\ &\mathbf{S}_{jk}\in\mathbb{C}^{\abs{\Phi_{jk}}\times\abs{\Phi_{jk}}}\,,~\mathbf{L}_{jk}\in\mathbb{H}^{\abs{\Phi_{jk}}\times\abs{\Phi_{jk}}}\,,~\forall (j,k) \in \mathcal{E}\,; \nonumber
    \\
    \text{s.t.}~~
    &\sum_{(j,k)\in\mathcal{E}}\operatorname{diag}(\mathbf{S}_{jk})^{\Phi_{j}} = \nonumber \\& \hspace{0.7cm}\sum_{(i,j)\in\mathcal{E}}\operatorname{diag}\left(\mathbf{S}_{ij}-\mathbf{Z}_{ij} \mathbf{L}_{ij}\right) +\mathbf{s}_j^G - \mathbf{s}_j^L\,,~~\forall j \in \mathcal{N}\,, \label{eq:net_pow_bal_sdp}\\
    &\mathbf{V}_{\!j} = \mathbf{V}_{\!i}^{\Phi_{ij}} - (\mathbf{Z}_{ij} \mathbf{S}_{ij}^H  + \mathbf{S}_{ij}\mathbf{Z}_{ij}^H)+\mathbf{Z}_{ij} \mathbf{L}_{ij} \mathbf{Z}_{ij}^H\,, \nonumber \\ &\hspace{4.5cm}\forall (i,j) \in \mathcal{E}\,, \label{eq:ohms_law_sdp}\\
    &\bmat{\mathbf{V}_{\!i}^{\Phi_{ij}} & \mathbf{S}_{ij} \\ \mathbf{S}_{ij}^H & \mathbf{L}_{ij}} \succeq 0\,,~~\forall (i,j) \in \mathcal{E}\,, \label{eq:net_semidef_sdp} \\ 
    &\operatorname{rank}\left(\bmat{\mathbf{V}_{\!i}^{\Phi_{ij}} & \mathbf{S}_{ij} \\ \mathbf{S}_{ij}^H & \mathbf{L}_{ij}}\right) = 1\,,~~\forall (i,j) \in \mathcal{E}\,, \label{eq:rank_1_net_sdp} \\
    &\mathbf{V}_{\!0} = \abs{V_0}\, \boldsymbol{\Gamma}\,, \label{eq:v0_sdp}\\
     &(\underline{\mathbf{v}}_{j})^2 \le \operatorname{diag}(\mathbf{V}_{\!j}) \le (\overline{\mathbf{v}}_{\!j})^2 \,, ~~\forall j \in \mathcal{N}\,, \label{eq:volt_con_sdp}\\
     &\operatorname{diag}(\mathbf{L}_{ij}) \le (\overline{\boldsymbol\ell}_{ij})^2 \,, ~~\forall (i,j) \in \mathcal{E}\,, \label{eq:curr_con_sdp}\\
    &\underline{\mathbf{p}}_{j}^G \le \operatorname{Re}\{\mathbf{s}_j^G\} \le \overline{\mathbf{p}}_{j}^G\,,~~\forall j \in \mathcal{N}\setminus\{0\}\,, \label{eq:der_plim_sdp}\\
    &\underline{\mathbf{q}}_{j}^G \le \operatorname{Imag}\{\mathbf{s}_j^G\} \le \overline{\mathbf{q}}_{j}^G\,,~~\forall j \in \mathcal{N}\setminus\{0\}\,, \label{eq:der_qlim_sdp}\\
     &\abs{\mathbf{s}_j^G}^2 \le (\mathbf{s}_{j,\text{rated}}^G)^2\,,~~\forall j \in \mathcal{N}\setminus\{0\}\,, \label{eq:der_slim_sdp}
\end{align}
where
\begin{equation}
    \boldsymbol{\Gamma} = \bmat{1 & a & a^2 \\ a^2 & 1 & a \\ a & a^2 & 1}\,,~~ a = e^{\,\text{j} 2\pi/3}. \label{eq:Gamma_bal}
\end{equation}

The objective of the OPF~\eqref{eq:encap_SDP} is to minimize the cost function $C(\cdot)$.
This cost function could represent various distribution system objectives, such as network loss minimization, voltage deviations from nominal, conservation voltage reduction, or PV hosting capacity.
In this paper, we choose to minimize the real power supplied to the feeder at the substation.
That is,
\begin{equation}
\label{eq:cost_fun}
    C(\mathbf{s}_{i}^G) = \mathbf{1}^\top \operatorname{Re}\{\mathbf{s}_0^G\}\,,
\end{equation}
where the (column) vector $\mathbf{s}_i^G$ denotes the power injected by distributed generation (or the substation) in each phase of node~$i$.
We assume, without loss of generality, that the substation is node $0$.
The 3-by-1 vector of ones, denoted by~$\mathbf{1}$, in~\eqref{eq:cost_fun} sums the real power injected into the feeder at the substation across all three phases.
We assume that real power can flow in reverse from the distribution grid to the transmission grid (i.e., $C$ can be negative).
This choice of objective function is advantageous because it tends to lead to i) reduced network losses, ii) conservation voltage reduction (when voltage-dependent loads are modeled), and iii) minimal curtailment of renewable-based DER generation.

We use the BFM to model the nonlinear power flow equations in ~\eqref{eq:net_pow_bal_sdp}--\eqref{eq:rank_1_net_sdp}~\cite{Low2014}.
The power balance constraint~\eqref{eq:net_pow_bal_sdp} ensures that all complex power leaving node $j$ in outgoing branches is equal to the complex power entering node $j$ from incoming branches plus the net injection (generation minus load) at node $j$.
For a branch $(j,k) \in \mathcal{E}$, the matrix $\mathbf{S}_{jk}$ denotes the complex power flowing into the branch from node $j$, $\mathbf{Z}_{jk}$ is the phase impedance matrix, and $\mathbf{L}_{jk}$ is a Hermitian matrix, whose diagonal is the current magnitudes squared in each phase.
The total complex power consumption of the load at node $j$ is denoted by $\mathbf{s}_j^L$.
The superscript $\Phi$ denotes a projection into the specified phases~\cite{Low2014}.
The constraint~\eqref{eq:ohms_law_sdp} captures the voltage drop across each branch, where the diagonal of the Hermitian matrix $\mathbf{V}_{\!j}$ is the voltage magnitudes squared in each phase of node $j$.
Finally, the positive semi-definite and rank constraints in~\eqref{eq:net_semidef_sdp}--\eqref{eq:rank_1_net_sdp} arise from the definition of slack variables in the BFM derivation (see~\cite{Low2014} for details).

We model the load $\mathbf{s}_j^L$ as a combination of constant power loads and constant impedance loads.
That is,
\begin{equation}
\label{eq:load_model}
    \mathbf{s}_j^L = \mathbf{s}_{j}^\text{PQ} + \operatorname{diag}\left(\mathbf{V}_{\!j} (\mathbf{Y}_j^{L})^*\right)\,,
\end{equation}
where $\mathbf{s}_{j}^\text{PQ}$ is the constant power part of the load at node $j$, and $\mathbf{Y}_j^{L}$ is the (diagonal) matrix of load admittances.
For constant power loads, the value of $\mathbf{s}_{j}^\text{PQ}$ is based on an average forecast of the load.
For constant impedance loads, $\mathbf{Y}_j^{L}$ may either be a forecasted value (e.g., representing voltage dependent loads inside a building), or it could be fixed (e.g., representing transformer core losses or the shunt capacitance of distribution cables).

Equation~\eqref{eq:v0_sdp} forces the voltages at the substation to be balanced, but allows the voltage magnitude $\abs{V_0}$ to vary.\footnote{It is assumed that the voltage magnitude at the feeder head can be controlled through the adjustment of substation voltage regulator setpoints.}
Constraints on the voltage magnitudes at node $j$ are enforced by~\eqref{eq:volt_con_sdp}, where $\underline{\mathbf{v}}_{j}\,,\overline{\mathbf{v}}_{j} \in \mathbb{R}^{\abs{\Phi_j}}$ are the vectors of voltage limits on each phase, and the square is applied element-wise.
Similarly,~\eqref{eq:curr_con_sdp} imposes an upper limit on branch current magnitudes.

Finally, equations~\eqref{eq:der_plim_sdp} and~\eqref{eq:der_qlim_sdp} represent rectangular constraints on the real and reactive power output of the DER, and~\eqref{eq:der_slim_sdp} enforces a circular apparent power rating.
For DER, we assume $\underline{\mathbf{p}}_{j}^G = \mathbf{0}$ and $\overline{\mathbf{p}}_{j}^G$ is determined by an average forecast of the maximum available power output of the dc source.
Furthermore, the reactive power constraints for DER are typically at least $\pm44\%$ of the apparent power rating, per the IEEE 1547 standard~\cite{IEEE1547}.
Also, note that we leave the complex power injection at the feeder head ($\mathbf{s}_{0}^G$) unconstrained.

\subsection{Convexity and Uniqueness of the Three-Phase OPF}

The optimization problem~\eqref{eq:encap_SDP} is convex, other than the rank constraints~\eqref{eq:rank_1_net_sdp}.
If we relax~\eqref{eq:rank_1_net_sdp} (by removing it), then the above three-phase OPF is in the form of a semi-definite program (SDP), which can be solved efficiently using commercially available solvers.
After solving the relaxed problem, if the constraints~\eqref{eq:rank_1_net_sdp} are met, then we have found a globally optimal solution to the original unrelaxed problem (i.e., the relaxation is exact or tight).
If the constraints~\eqref{eq:rank_1_net_sdp} are not satisfied, then the solution of the SDP provides a lower bound on the global optimal solution, but is not a feasible power flow solution.
In this case, several techniques for finding feasible solutions that are near globally optimal have been proposed~\cite{Molzahn2017}.
Sufficient conditions for the exactness of the SDP relaxation for the BFM-based multi-phase OPF are explored in~\cite{Low2014exactness}.
For many radial distribution feeders, the BFM-based SDP relaxation has been shown to be exact (to within numerical precision)~\cite{Low2014}.

Finally, note that the BFM uses the voltage and current magnitudes squared and complex power flows on each branch as variables.
However, if the solution of the SDP relaxation satisfies the rank constraints~\eqref{eq:rank_1_net_sdp}, then the unique three-phase voltage and current phasors in the network can be recovered from the BFM variables (see Lemma 2 and Algorithm 2 of~\cite{Low2014}).

\section{Handling Forecasting Uncertainty}
\label{sec:jacobian}
\begin{figure}
    \centering
    \begin{tikzpicture}
    \node[inner sep=0pt] (n1) at (-5,0)    {\includegraphics[width=0.48\textwidth,trim=0 0 0 0, clip=true]{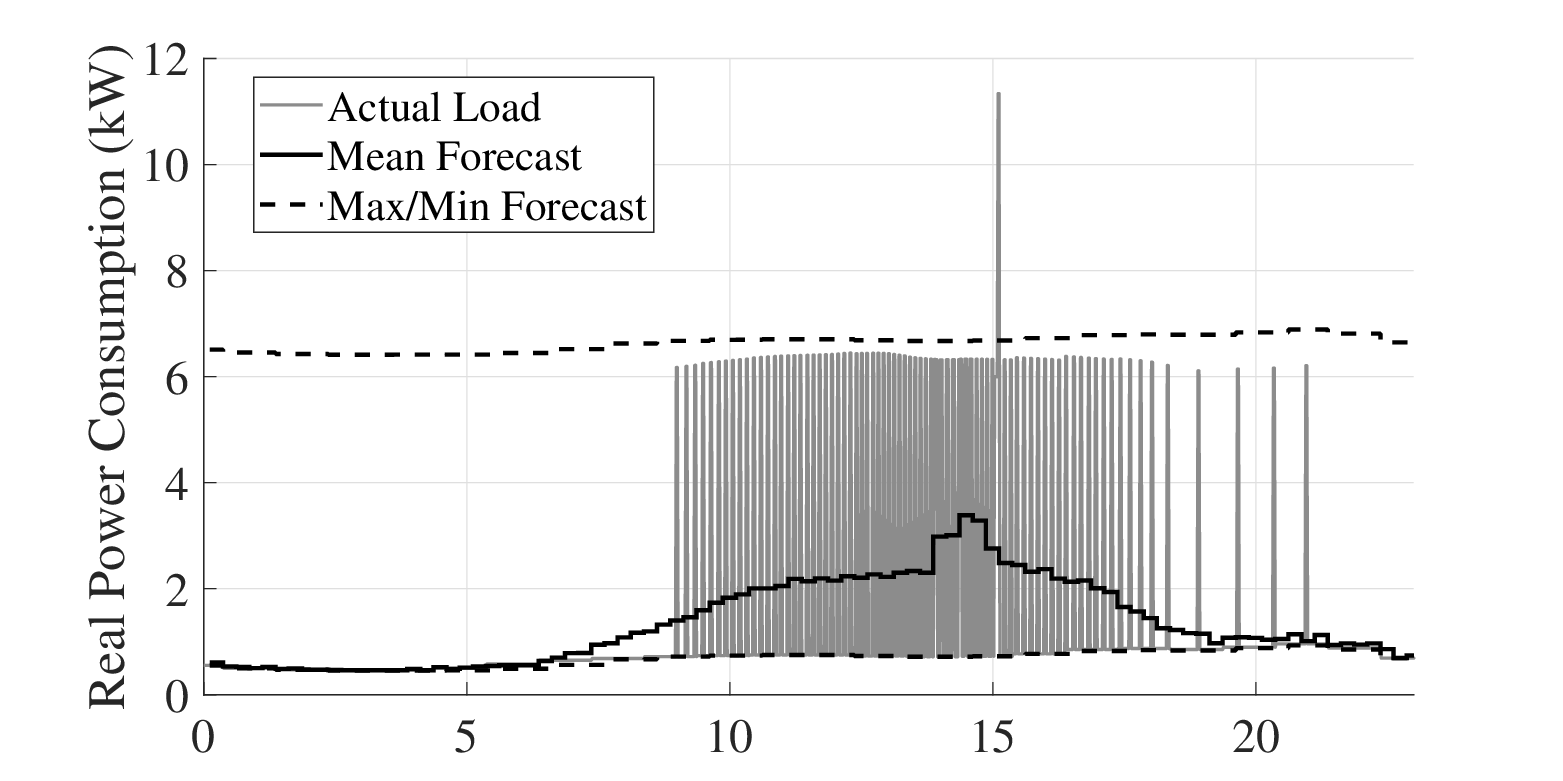}};
    \node[inner sep=0pt] (n1) at (-5,-4.5)    {\includegraphics[width=0.48\textwidth,trim=0 0 0 0, clip=true]{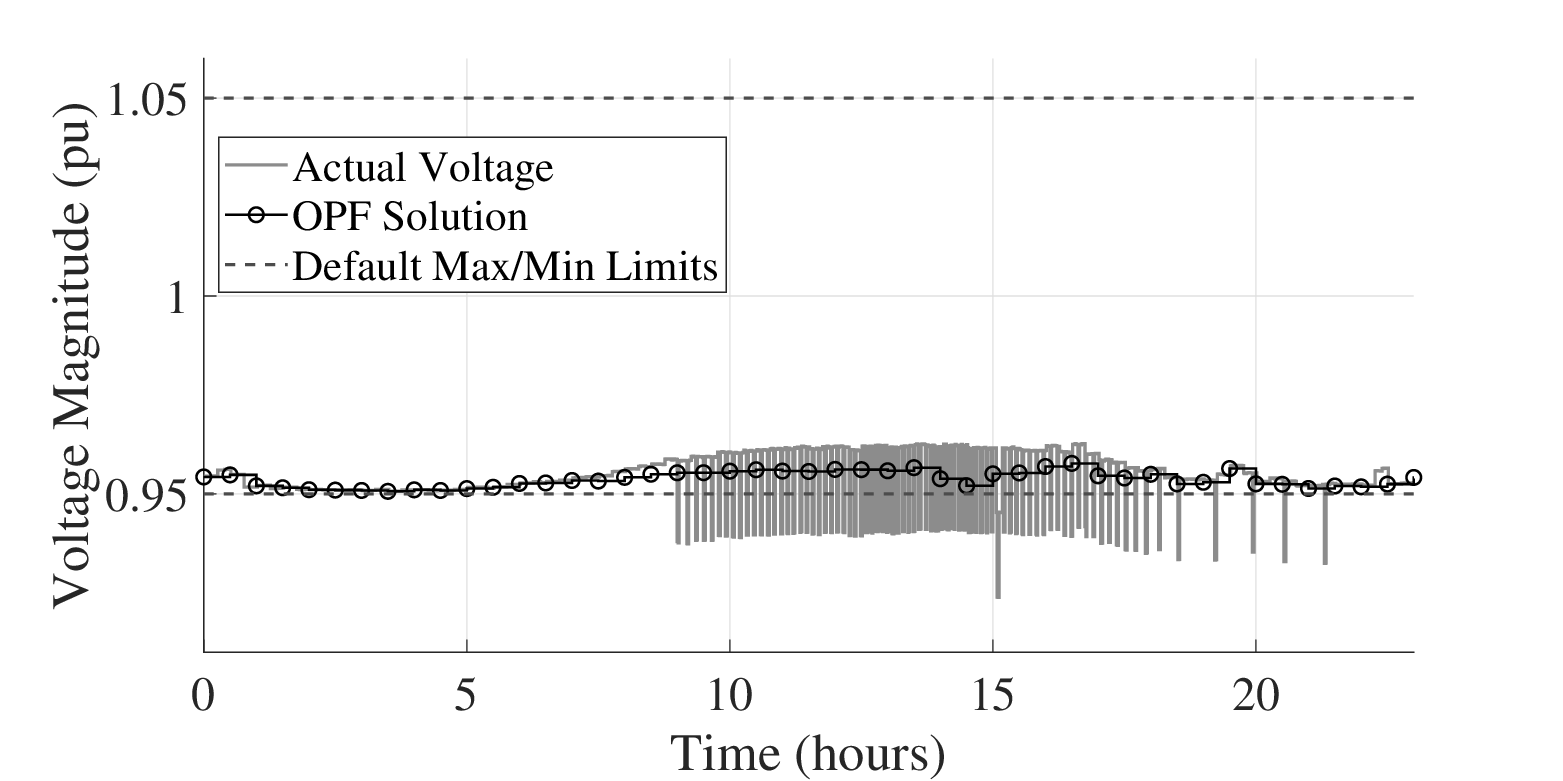}};
    \node[inner sep=0pt] (n1) at (-9.5,0)
    {\normalsize (a)};
    \node[] (n2) at (-9.5,-4.5) {\normalsize (b)};
    \end{tikzpicture}
    \caption{Real power consumption and voltage magnitude for an individual single-family home over a 24-hour period. (a) Comparison of actual power consumption against 15-minute average, minimum and maximum forecasts. (b) Comparison of actual voltage magnitudes and solution of OPF with constraints based on average load forecasts.}
    \label{fig:load_forecast}
\end{figure}

In the above OPF formulation, the load and DER generation at each node is modeled as a constant over a given window in time (e.g., 15 minutes), based on an average forecast.
However, as seen in Fig.~\ref{fig:load_forecast}(a), the actual power consumption of a building can vary drastically from the forecasted average due to the switching of loads.
Thus, if the voltage constraints at low-voltage nodes in the OPF are enforced based on average load forecasts, then voltage violations can occur when loads switch on and off.
This can be seen around time $t=15$ hours in Fig.~\ref{fig:load_forecast}(b).
When the air conditioner switches on, the voltage drops below the lower limit even though the OPF solution (based on the average load forecast) predicts that no voltage violations will occur.
Similar voltage violations can occur when available generation from renewable resources vary from their average forecasts.

To address this issue, we propose to set the voltage constraints based on worst-case load and DER generation forecasts.
We begin by assuming that, in addition to the mean 15-minute load and generation forecast for each building, we also obtain a 15-minute forecast of the (worst-case) minimum and maximum expected load and DER generation (see Fig.~\ref{fig:load_forecast}).
It is envisioned that these forecasts could be obtained from historical AMI (i.e., smart meter) data.
From the mean, minimum, and maximum forecasts, we can calculate the largest expected deviations (positive and negative) in the real and reactive power injection at node $\ell$.
That is,
\begin{align}
    \label{Eq:delta_power1}
    \Delta \mathbf{p}_{\ell}^- &=  (\hat{\mathbf{p}}_\ell^L-\overline{\mathbf{p}}_\ell^L)-(\hat{\mathbf{p}}_\ell^G-\underline{\mathbf{p}}_\ell^G)\,, \\
    \Delta \mathbf{p}_{\ell}^+ &=  (\hat{\mathbf{p}}_\ell^L-\underline{\mathbf{p}}_\ell^L)-(\hat{\mathbf{p}}_\ell^G-\overline{\mathbf{p}}_\ell^G)\,, \\
    \Delta \mathbf{q}_{\ell}^- &=  (\hat{\mathbf{q}}_\ell^L-\overline{\mathbf{q}}_\ell^L)-(\hat{\mathbf{q}}_\ell^G-\underline{\mathbf{q}}_\ell^G)\,, \\
    \Delta \mathbf{q}_{\ell}^+ &=  (\hat{\mathbf{q}}_\ell^L-\underline{\mathbf{q}}_\ell^L)-(\hat{\mathbf{q}}_\ell^G-\overline{\mathbf{q}}_\ell^G)\,,
     \label{Eq:delta_power2}
\end{align}
where $\hat{\mathbf{p}}_\ell^L\,,\hat{\mathbf{q}}_\ell^L\,,\hat{\mathbf{p}}_\ell^G\,,\hat{\mathbf{q}}_\ell^G \in \mathbb{R}^{\abs{\Phi_\ell}}$ are the mean real and reactive power consumption forecasts for the load and DER generation at node $\ell$, 
$\overline{\mathbf{p}}_\ell^L\,,\overline{\mathbf{q}}_\ell^L\,,\overline{\mathbf{p}}_\ell^G\,,\overline{\mathbf{q}}_\ell^G \in \mathbb{R}^{\abs{\Phi_\ell}}$ are the maximum forecasts, and $\underline{\mathbf{p}}_\ell^L\,,\underline{\mathbf{q}}_\ell^L\,,\underline{\mathbf{p}}_\ell^G\,,\underline{\mathbf{q}}_\ell^G \in \mathbb{R}^{\abs{\Phi_\ell}}$ are the minimum forecasts. 
The largest expected increases in real and reactive power injection at node $\ell$ are denoted by $\Delta \mathbf{p}_{\ell}^+\,,\Delta \mathbf{q}_{\ell}^+ \ge 0$, and the largest expected decreases in real and reactive power injection are 
$\Delta \mathbf{p}_{\ell}^-\,,\Delta \mathbf{q}_{\ell}^- \le 0$.

Next, we find the sensitivity of the voltage magnitude at node~$j$ to changes in power injection at node~$\ell$.
Linearizing the three-phase power flow equations around a given operating point, we obtain\footnote{We choose to linearize about the operating point obtained from solving a three-phase power flow, where we fix the voltage magnitude at the feeder head to 1.0 pu voltage, and assume all DER output the maximum available real power at unity power factor. Our case study results illustrate that this approximation is acceptable.}
\begin{equation}
        \bmat{\Delta \mathbf{v} \\ \Delta \boldsymbol{\theta}} = \mathbf{J}^{-1} \bmat{\Delta \mathbf{p} \\ \Delta \mathbf{q}}\,, \\
\end{equation}
where the vectors  $\Delta \mathbf{v}$, $\Delta \boldsymbol\theta$, $\Delta \mathbf{p}$, and $\Delta \mathbf{q}$ collect voltage magnitudes, voltage angles, real power injections, and reactive power injections over all $N$ nodes (with $N=\abs{\mathcal{N}}$).
In particular, $\Delta \mathbf{v}_j \in \mathbb{R}^{\abs{\Phi_j}}$ is the vector of deviations in voltage magnitudes at node~$j$, 
and $\Delta \mathbf{p}_j\,, \Delta \mathbf{q}_j \in \mathbb{R}^{\abs{\Phi_j}}$ are the vectors of deviations in real and reactive power injection, respectively.
The inverse of the three-phase power flow Jacobian, $\mathbf{J}^{-1}$, is a full matrix that contains the sensitivities of the voltage magnitudes with respect to real and reactive power injections, stored in submatrices $\pd{\mathbf{v}_j}{\mathbf{p}_\ell}$ and $\pd{\mathbf{v}_j}{\mathbf{q}_\ell}$, respectively~\cite{birt1975}.

Hence, the largest expected change (both positive and negative) in the voltage magnitude in phase~$\phi$ of node~$j$ due to a change in real power injection in phase~$\psi$ of node~$\ell$ is
\begin{align}
    \Delta V_{j\phi,\ell\psi}^{P+} = 
    \begin{cases}
        \alpha_{j\phi,\ell\psi} \Delta P_{\ell\psi}^+\,, & \alpha_{j\phi,\ell\psi} \ge 0\\
        \alpha_{j\phi,\ell\psi} \Delta P_{\ell\psi}^-\,, & \alpha_{j\phi,\ell\psi} < 0\\
    \end{cases}\,, \\
    \Delta V_{j\phi,\ell\psi}^{P-} = 
    \begin{cases}
        \alpha_{j\phi,\ell\psi} \Delta P_{\ell\psi}^-\,, & \alpha_{j\phi,\ell\psi} \ge 0 \\
        \alpha_{j\phi,\ell\psi} \Delta P_{\ell\psi}^+\,, & \alpha_{j\phi,\ell\psi} < 0 
    \end{cases} \,,
\end{align}
where $\alpha_{j\phi,\ell\psi}$ is the element of $\mathbf{J}^{-1}$ corresponding to row $\phi$ and column $\psi$ of $\pd{\mathbf{v}_j}{\mathbf{p}_\ell}$, and $\Delta P_{\ell\psi}^+\,, \Delta P_{\ell\psi}^-$ are the elements of $\Delta \mathbf{p}_{\ell}^+\,,\Delta \mathbf{p}_{\ell}^-$ corresponding to phase $\psi$.
Similarly, for changes in reactive power injection, we have
\begin{align}
    \Delta V_{j\phi,\ell\psi}^{Q+} = 
    \begin{cases}
        \beta_{j\phi,\ell\psi} \Delta Q_{\ell\psi}^+\,, & \beta_{j\phi,\ell\psi} \ge 0\\
        \beta_{j\phi,\ell\psi} \Delta Q_{\ell\psi}^-\,, & \beta_{j\phi,\ell\psi} < 0\\
    \end{cases}\,, \\
    \Delta V_{j\phi,\ell\psi}^{Q-} = 
    \begin{cases}
        \beta_{j\phi,\ell\psi} \Delta Q_{\ell\psi}^-\,, & \beta_{j\phi,\ell\psi} \ge 0 \\
        \beta_{j\phi,\ell\psi} \Delta Q_{\ell\psi}^+\,, & \beta_{j\phi,\ell\psi} < 0 
    \end{cases} \,,
\end{align}
where $\beta_{j\phi,\ell\psi}$ is the element of $\mathbf{J}^{-1}$ corresponding to row $\phi$ and column $\psi$ of $\pd{\mathbf{v}_j}{\mathbf{q}_\ell}$, and $\Delta Q_{\ell\psi}^+\,, \Delta Q_{\ell\psi}^-$ are the elements of $\Delta \mathbf{q}_{\ell}^+\,,\Delta \mathbf{q}_{\ell}^-$ corresponding to phase $\psi$.\footnote{Note that $\Delta V_{j\phi,\ell\psi}^{P+}\,, \Delta V_{j\phi,\ell\psi}^{Q+}\ge0$ and $\Delta V_{j\phi,\ell\psi}^{P-}\,,\Delta V_{j\phi,\ell\psi}^{Q-} \le 0$.}

To find the largest possible increase (decrease) in voltage magnitude in phase $\phi$ of node~$j$, we could simply take the sum of $\Delta V_{j\phi,\ell\psi}^{P+}$ and $\Delta V_{j\phi,\ell\psi}^{Q+}$ ($\Delta V_{j\phi,\ell\psi}^{P-}$ and $\Delta V_{j\phi,\ell\psi}^{Q-}$) for all phases~$\psi$ of all nodes~$\ell$.
However, this would correspond to all of the load and available renewable generation in the feeder simultaneously changing to their maximum or minimum forecasted value (depending of the sign of the Jacobian coefficients).
The probability of this happening is extremely small, and this approach would be overly conservative.
Instead, we find the change in voltage that would result from the $\kappa$ simultaneous changes in power injection that would have the largest contribution to the change in voltage.
That is,
\begin{align}
    \Delta V_{j\phi}^+ &= \operatorname{summaxk}\left(\{\Delta V_{j\phi,\ell\psi}^{P+},\Delta V_{j\phi,\ell\psi}^{Q+}\}_{\psi \in \Phi_\ell\,,~\ell \in \mathcal{N}}\right)\,, \\
    \Delta V_{j\phi}^- &= \operatorname{summaxk}\left(\{\Delta V_{j\phi,\ell\psi}^{P-},\Delta V_{j\phi,\ell\psi}^{Q-}\}_{\psi \in \Phi_\ell\,,~\ell \in \mathcal{N}}\right)\,,
\end{align}
where the operator $\operatorname{summaxk}(\cdot)$ finds the sum of the largest (in magnitude) $\kappa$ elements of a set.\footnote{Note that $\Delta V_{j\phi}^+ \ge 0$ and $\Delta V_{j\phi}^- \le 0$.}
The choice of $\kappa$ allows the system operator to determine how conservative to be in accounting for voltage violations (i.e., a higher value of $\kappa$ will be more conservative, while a lower value of $\kappa$ may result in a lower value of the objective cost $C$ at the risk of more voltage violations.)
In the case study of Sec.~\ref{sec:results}, we select $\kappa=3$.

Finally, we modify the voltage constraints~\eqref{eq:volt_con_sdp} to obtain
\begin{equation}
\label{eq:mod_volt_con_sdp}
    (\underline{\mathbf{v}}_{j}-\Delta \mathbf{v}_j^-)^2 \le \operatorname{diag}(\mathbf{V}_{\!j}) \le (\overline{\mathbf{v}}_{\!j}-\Delta \mathbf{v}_j^+)^2 \,, ~~\forall j \in \mathcal{N}\,,
\end{equation}
where $\Delta \mathbf{v}_j^- = [\Delta V_{j\phi}^-]_{\phi \in \Phi_j}$ and $\Delta \mathbf{v}_j^+ = [\Delta V_{j\phi}^+]_{\phi \in \Phi_j}$.
Essentially,~\eqref{eq:mod_volt_con_sdp} introduces a conservative margin to the voltage constraints (by increasing the lower limit and decreasing the upper limit) based on the worst-case change in voltage we expect to occur due to uncertainty in the load and DER generation forecasts.

\section{Modeling Split-Phase Secondaries}
\label{sec:split_phase}

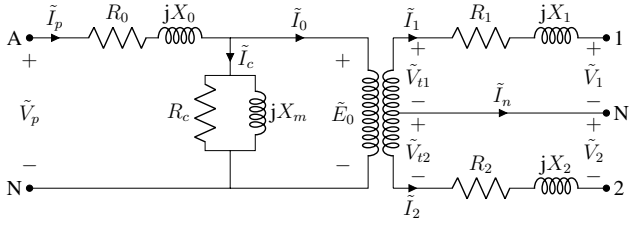
\begin{figure}
    \centering
    \resizebox{0.48\textwidth}{!}{
        \begin{circuitikz}[
american voltages,
longL/.style = {cute inductor, inductors/scale=1.0, inductors/width=1.2, inductors/coils=10}
]

\large
\draw

  (0.5,0) to [open, v<={$\tilde{V}_{p}$}] (0.5,3)
  to [short, i=$\tilde{I}_p$] (1.5,3)
  to [R, l=$R_{0}$] (3,3)
  to [L, l=$\text{j} X_{0}$] (4,3)
  to (4.5,3)
  to [short, i=$\tilde{I}_c$] (4.5,2.25)
  to (4,2.25)
  to [R, l_=$R_{c}$] (4,0.75)
  to (4.5,0.75)
  to (4.5,0)
  to (0.5,0)
  
  (4.5,2.25) to (5,2.25)
  to [L, l=$\text{j} X_m$] (5,0.75)
  to (4.5,0.75)

  (4.5,3) to [short, i=$\tilde{I}_0$] (7.25,3)
  to [longL] (7.25,0)
  to (4.5,0)
  (7.75,0) to [longL,name=L2] (7.75,3)

  (7.75,3) to [short, i=$\tilde{I}_1$] (8.5,3)
  to [R, l=$R_{1}$] (10.5,3) to [L, l=$\text{j} X_{1}$] (11.5,3) to (12,3)

  (7.75,0) to [short, i_=$\tilde{I}_2$] (8.5,0)
  to [R, l=$R_{2}$] (10.5,0) to [L, l=$\text{j} X_{2}$] (11.5,0) to (12,0)

  (L2.midtap) to [short, i=$\tilde{I}_n$] (12,1.5)

  (6.75,0) to [open, v<={$\tilde{E}_{0}$}] (6.75,3)
    
  (8.25,0) to [open, v<={$\tilde{V}_{t2}$}] (8.25,1.5)
  (8.25,1.5) to [open, v<={$\tilde{V}_{t1}$}] (8.25,3)

  (11.75,0) to [open, v<={$\tilde{V}_{2}$}] (11.75,1.5)
  (11.75,1.5) to [open, v<={$\tilde{V}_{1}$}] (11.75,3);

  \draw[fill] (0.5,0) circle (0.0625) node[left] {N};
  \draw[fill] (0.5,3) circle (0.0625) node[left] {A};

  \draw[fill] (12,0) circle (0.0625) node[right] {2};
  \draw[fill] (12,3) circle (0.0625) node[right] {1};
  \draw[fill] (12,1.5) circle (0.0625) node[right] {N};


    
  
  
\end{circuitikz}
    }
    \caption{Circuit diagram for split-phase, center-tapped distribution transformer model with core losses.}
    \label{fig:split_phase_xfrm_diag}
\end{figure}

\begin{figure}
    \centering
    \resizebox{0.28\textwidth}{!}{
        \begin{circuitikz}[
american voltages,
longL/.style = {cute inductor, inductors/scale=1.0, inductors/width=1.2, inductors/coils=10},
shortG/.style = {generic, resistors/scale=0.75}
]

\large
\draw

  (7.75,3) to [short, i=$\tilde{I}_{1,i\ell}$] (8.5,3)
  to [R, l=$R_{1,i\ell}$] (10.5,3) to [L, l=$\text{j} X_{1,i\ell}$] (11.5,3) to (12,3)

  (7.75,0) to [short, i_=$\tilde{I}_{2,i\ell}$] (8.5,0)
  to [R, l=$R_{2,i\ell}$] (10.5,0) to [L, l=$\text{j} X_{2,i\ell}$] (11.5,0) to (12,0)

  (L2.midtap) to [short, i=$\tilde{I}_{n,i\ell}$] (12,1.5)
    
  (8.25,0) to [open, v<={$\tilde{V}_{2,i}$}] (8.25,1.5)
  (8.25,1.5) to [open, v<={$\tilde{V}_{1,i}$}] (8.25,3)

  (11.75,0) to [open, v<={$\tilde{V}_{2,\ell}$}] (11.75,1.5)
  (11.75,1.5) to [open, v<={$\tilde{V}_{1,\ell}$}] (11.75,3)
  
  (12,3) to (12.5,3) to [shortG] (12.5,1.5) to (12,1.5)
  (12,1.5) to (12.5,1.5) to [shortG] (12.5,0) to (12,0)
  (12,3) to (13.5,3) to [shortG] (13.5,0) to (12,0)
  ;

  \draw[fill] (7.75,0) circle (0.0625) node[left] {2};
  \draw[fill] (7.75,3) circle (0.0625) node[left] {1};
  \draw[fill] (7.75,1.5) circle (0.0625) node[left] {N};

  \draw[fill] (12.5,0) circle (0.0625);
  \draw[fill] (12.5,3) circle (0.0625);
  \draw[fill] (12.5,1.5) circle (0.0625);


    
  
  
\end{circuitikz}
    }
    \caption{Circuit diagram for a split-phase secondary consisting of triplex cable and load.}
    \label{fig:split_phase_diag}
\end{figure}
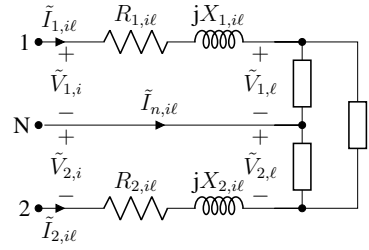

In this section, we describe how models of center-tapped distribution transformers and split-phase secondaries can be incorporated in the three-phase OPF\@.
A circuit diagram for a split-phase, center-tapped distribution transformer is shown in Fig.~\ref{fig:split_phase_xfrm_diag}~\cite{kersting2018}.
The primary winding of the transformer is connected phase-to-neutral, whereas the secondary winding of the transformer is center-tapped to provide two 120-V circuits.
The series impedances of the transformer coils are denoted by $Z_0 = R_0 + \text{j}X_0$, $Z_1 = R_1 + \text{j}X_1$, and $Z_2 = R_2 + \text{j}X_2$.
The core losses of the transformer are modeled by a resistance, $R_c$, in parallel with the magnetizing reactance, $X_m$.
Thus, the shunt impedance associated with the transformer core is
\begin{equation}
    Z_c = \frac{\text{j}R_c X_m}{R_c+\text{j}X_m}\,.
\end{equation}
The equations governing the ideal split-phase transformer in Fig.~\ref{fig:split_phase_xfrm_diag} are
\begin{gather}
\label{eq:E0}
    \tilde{E}_0 = n_t \tilde{V}_{t1} = n_t \tilde{V}_{t2}\,,\\
    \label{eq:I0}
    \tilde{I}_0 = \frac{1}{n_t}(\tilde{I}_1-\tilde{I}_2)\,, \\
    n_t = \frac{V_{bp,\ell n}}{V_{bs,\ell n}}\,,
\end{gather}
where $V_{bp,\ell n}$ and $ V_{bs,\ell n}$ are the base line-to-neutral voltages on the transformer primary and secondary, respectively.
The secondary windings of these transformers are typically connected to building loads through triplex cables, as shown in Fig.~\ref{fig:split_phase_diag}.
We denote the series impedances of the triplex cables connecting node $i$ to node $\ell$ by $Z_{1,i\ell} = R_{1,i\ell} +\text{j}X_{1,i\ell}$ and $Z_{2,i\ell} = R_{2,i\ell} +\text{j}X_{2,i\ell}$.
Loads can be connected to these split-phase secondaries in a line-neutral configuration to provide 120-V service, or in a line-to-line configuration for 240-V service (see Fig.~\ref{fig:split_phase_diag}).

One approach for modeling these split-phase secondaries in a three-phase OPF formulation would be to include them as two-phase branches (i.e., one phase for each 120-V circuit).
However, there are a few issues with this approach.
First, under this approach, loads that are connected line-to-line (i.e., 240-V loads) essentially create a delta-connection with loads connected line-to-neutral (see Fig.~\ref{fig:split_phase_diag}).
It has been shown that these delta-connected loads tend to render the SDP relaxation of the BFM-based OPF inexact~\cite{Zhou2021}.
Second, it is envisioned that the load forecasts used in the OPF would be based on historical smart meter data; however, smart meters typically record only the total power consumed by a building rather than the power consumed on each phase.
Thus, forecasts for the load in each phase may not be available.

Instead, we model the split-phase secondaries by making two simplifying assumptions:\footnote{In practice, the first assumption is typically close to true based on the symmetrical construction of center-tapped transformers and triplex cables. The second assumption, however, is usually false as residential consumers typically don't actively balance loads between the 120-V circuits. Regardless, this approximation may be good enough if load imbalances in the split-phase secondaries do not significantly impact system voltages. To validate this, we simulate unbalanced loads in the numerical case studies of Sec.~\ref{sec:results}.}
\begin{enumerate}
    \item The series impedance of the transformer windings and triplex cables in each 120-V circuit are identical (i.e., $Z_1 = Z_2$ and $Z_{1,i\ell} = Z_{2,i\ell}$).
    \item The loads connected phase-to-neutral in each 120-V circuit are equal.
\end{enumerate}
Under these assumptions, zero current will flow in the neutral wire of the split-phase secondary (i.e., $\tilde{I}_1 = -\tilde{I}_2$, $\tilde{V}_1 = \tilde{V}_2$), and the split-phase secondary can be modeled as a single-phase circuit connected line-to-line.
Furthermore, note that~\eqref{eq:I0} simplifies to 
\begin{equation}
    \tilde{I}_0 = \frac{2}{n_t}\tilde{I}_1\,.
\end{equation}

\begin{figure}
    \centering
    \resizebox{0.48\textwidth}{!}{
        \begin{circuitikz}[
american voltages,
longL/.style = {cute inductor, inductors/scale=1.0, inductors/width=1.2, inductors/coils=10},
shortG/.style = {generic, resistors/scale=0.75}
]

\large
\draw

  (0.5,0) to [open, v<={$\tilde{V}_{p,\text{pu}}$}] (0.5,2)
  to [short, i=$\tilde{I}_{p,\text{pu}}$] (1.5,2)
  to [shortG, l=$Z_{0,\text{pu}}$] (3,2)
  to (3.5,2)
  to [shortG,l=$Z_{c,\text{pu}}$] (3.5,0)

  (0.5,0) to [short] (8.5,0)

  (3.5,2) to [short] (4,2)
  to [shortG, l=$Z_{1,\text{pu}}$] (5.5,2)
  to (6,2)
  to  [short] (6.5,2)
  to [shortG, l=$Z_{1,i\ell,\text{pu}}$] (8,2)
  to [short] (8.5,2)

  (2.9,0) to [open, v<={$\tilde{E}_{0,\text{pu}}$}] (2.9,2)

  (8.5,2) to [short] (9.125,2)
  to [shortG] (9.125,0)
  to (8.5,0)
  
  (6,0) to [open, v<={$\tilde{V}_{1,i,\text{pu}}$}] (6,2)
  (8.4,0) to [open, v<={$\tilde{V}_{1,\ell,\text{pu}}$}] (8.4,2);

  \draw[fill] (0.5,2) circle (0.0625);
  \draw[fill] (0.5,0) circle (0.0625);
  \draw[fill] (3.5,2) circle (0.0625);
  \draw[fill] (6,2) circle (0.0625);
  \draw[fill] (8.5,2) circle (0.0625);


    
  
  
\end{circuitikz}
    }
    \caption{Circuit diagram for T-equivalent split-phase, center-tapped distribution transformer model including core losses and triplex cable (in per unit).}
    \label{fig:split_phase_T}
\end{figure}
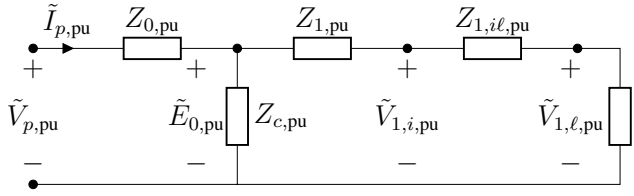

Next, we select base quantities
\begin{gather}
    I_{bp} = \frac{S_b}{3 V_{bp,\ell n}}\,,~~~Z_{bp} = \frac{3 (V_{bp,\ell n})^2}{S_b}\,, \\
    I_{bs} = \frac{1}{2}\frac{S_b}{3 V_{bs,\ell n}}\,,~~~Z_{bs} = 2 \frac{3 (V_{bs,\ell n})^2}{S_b}\,,
\end{gather}
where $I_{bp}$ and $Z_{bp}$ are the base current and impedance on the primary side, $I_{bs}$ and $Z_{bs}$ are the base current and impedance on the secondary side, and $S_b$ is the three-phase base apparent power.
Thus, in per unit, we obtain a T-equivalent circuit for the split-phase, center-tapped transformer, as shown in Fig.~\ref{fig:split_phase_T}.
Note that the per-unit series impedances $Z_{0,\text{pu}}$, $Z_{1,\text{pu}}$, and $Z_{1,i\ell,\text{pu}}$ can be incorporated as single-phase branches in the BFM\@.
Additionally, the shunt impedance $Z_{c,\text{pu}}$, which models transformer core losses, can be added as a constant impedance load by applying~\eqref{eq:load_model} at the artificially introduced node corresponding to $\tilde{E}_{0,\text{pu}}$ in Fig.~\ref{fig:split_phase_T}. 

\section{Numerical Case Study}
\label{sec:results}

\begin{figure}
    \begin{tikzpicture}
    \node[inner sep=0pt] (n1) at (-5,0)
    {\includegraphics[width=0.45\textwidth,trim=4cm 4cm 4cm 3cm, clip=true]{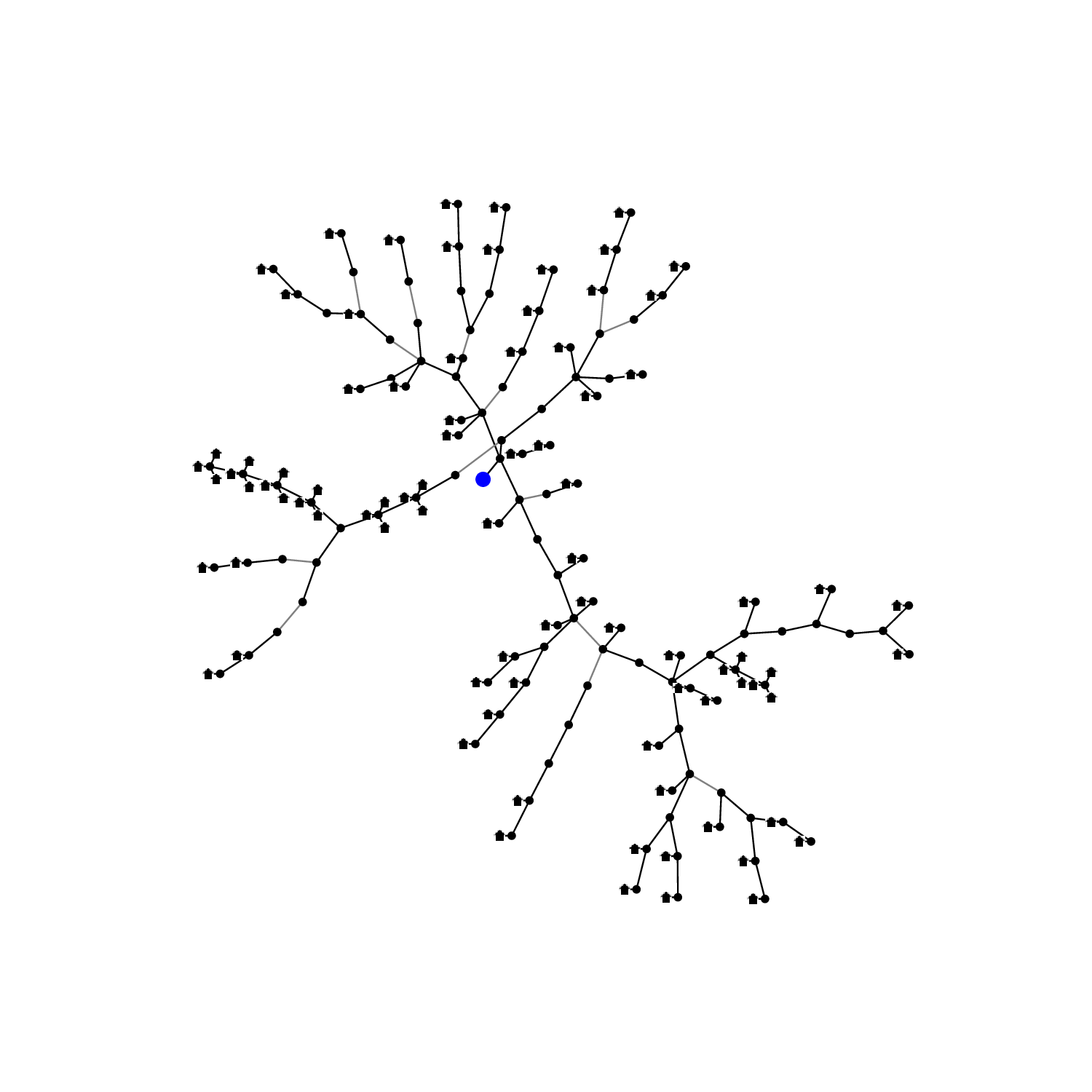}};
    \node[] (n2) at (-6.2,-0.7) {\scriptsize Substation};
    \draw [-stealth](-6.2,-0.5) -- (-5.9,0.2);
    \draw [stealth-](-6.2+0.1,-0.5+4.1) -- (-6.2+0.1,-0.2+4.1);
    \node[] (n2) at (-5.9+0.1,0+4.1) {\scriptsize Split-Phase Transformer + Triplex Lines + (Multiple) Residential Loads};
    \draw [stealth-](-6.2+0.3,-0.5+4.1) -- (-5.7+0.3,-0.4+4.1);
    \node[] (n2) at (-4.3+0.1,-0.35+4.1) {\scriptsize Medium-Voltage Node};

    \end{tikzpicture}
    \caption{ Modified section of R2-12.47-3 PNNL taxonomy feeder. Note that only medium-voltage lines are shown; each house in this diagram represents a center-tapped transformer and split-phase secondary (potentially feeding multiple homes).}
    \label{fig:taxonomy_feeder}
\end{figure}
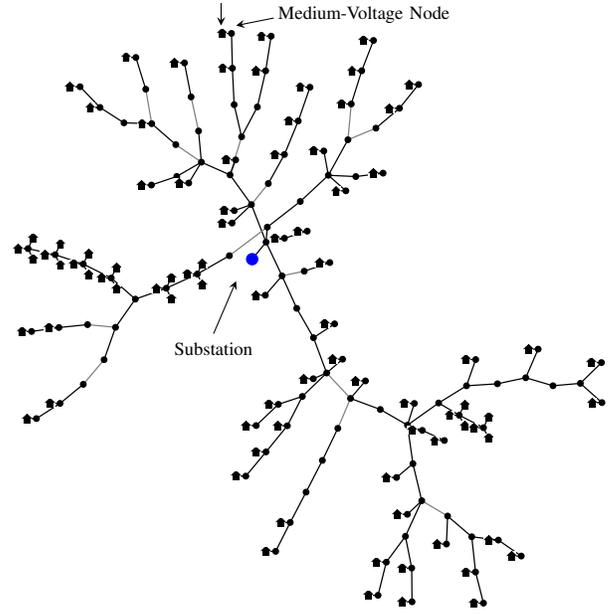

To validate the proposed approach, we use a modified section of the R2-12.47-3 taxonomy feeder from Pacific Northwest National Laboratory (PNNL), as shown in Fig.~\ref{fig:taxonomy_feeder}~\cite{schneider2008modern}. 
This 320-node feeder consists of primarily single-family homes that are connected to the medium-voltage distribution network through split-phase, center-tapped transformers and triplex cables. 
Each of the 230 residential loads are modeled as a mix of thermostatically-controlled devices (e.g., air conditioners and water heaters) and end-use loads (lighting, appliances, pool pumps, etc.). 
The feeder has also been randomly populated with rooftop photovoltaic solar up to a penetration level of 50\%.\footnote{Here, the solar penetration level is defined as the percentage of buildings that have rooftop solar installed, rather than a percentage of the peak feeder load. Thus, a 50\% penetration level means that roughly one out of every two households have rooftop solar.} 
Each solar panel is connected to the grid through an inverter operating in a constant PQ mode.
We assume that the PQ setpoints of the inverter can be adjusted every 15 minutes through the IEEE 1547 standard~\cite{IEEE1547}, based on the OPF solution.
The minute-by-minute power flows in the feeder are simulated based on weather data for a hot July day in Chicago, IL, USA\@.
The distribution system simulation software, GridLAB-D v4.3, is used to conduct these simulations~\cite{gridlabD}.

The proposed method for dynamically adjusting the voltage constraints in the OPF is implemented in Matlab R2022b, and the SDP relaxation of the BFM-based OPF is solved using the SeDuMi solver in CVX~\cite{cvx}.\footnote{All calculations and simulations are performed on a Dell Precision 7820 computer with two 2.20~GHz, 48-core Intel\textsuperscript{\textregistered} Xeon\textsuperscript{\textregistered} Gold 5220R CPUs.}
We leverage the Matlab link functionality in GridLAB-D to interface the OPF solution with the simulation of the distribution feeder.
More specifically, after every 15 minutes of simulation, GridLAB-D stops its calculations and calls a Matlab function to solve the OPF\@.
Based on the solution of the OPF, the inverter PQ setpoints and substation voltage regulator setpoints are adjusted, and GridLAB-D resumes simulation.

\subsection{Generating Forecast Data}
\label{sec:forecasts}
The proposed approach for improving the robustness of the OPF to uncertainty relies on 15-minute average, minimum, and maximum forecasts for the load and DER generation at each individual house.
To generate the forecasts for this study, we use GridLAB-D to simulate the feeder over an entire month, and record smart meter readings for each building.
This synthetic historical data is then used to build the required 15-minute forecasts.\footnote{A number of statistical methods could be used for generating these forecasts from the synthetic data. An investigation of which methods work best for performing this forecasting is outside the scope of this work.}
We emphasize that the forecasted load and solar generation profiles are different for each house.

\begin{figure}
    \begin{tikzpicture}
    \node[inner sep=0pt] (n1) at (-5,0)    {\includegraphics[width=0.45\textwidth,trim=0 0 0 0, clip=true]{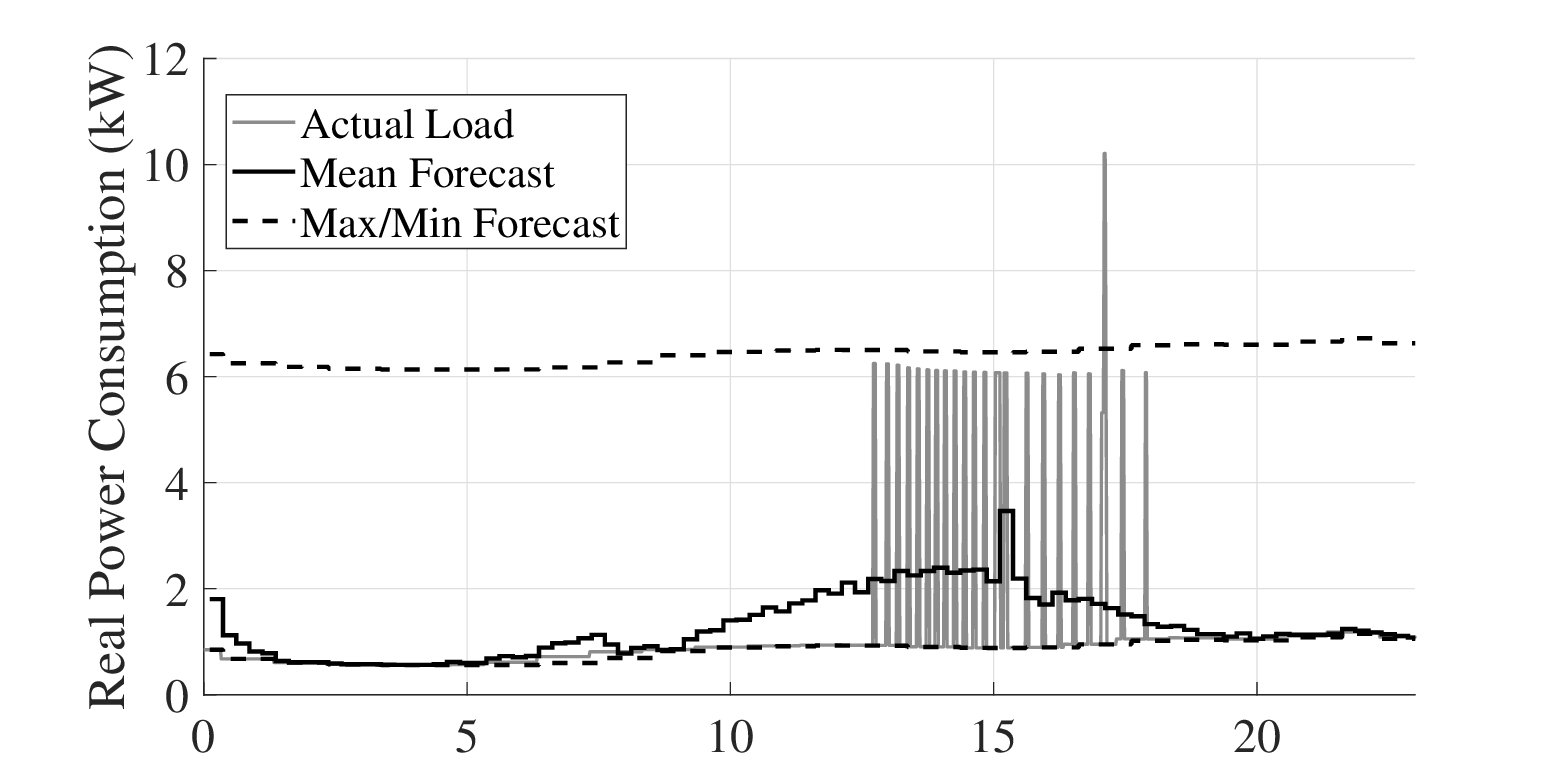}};
    \node[inner sep=0pt] (n1) at (-5,-4)    {\includegraphics[width=0.45\textwidth,trim=0 0 0 0, clip=true]{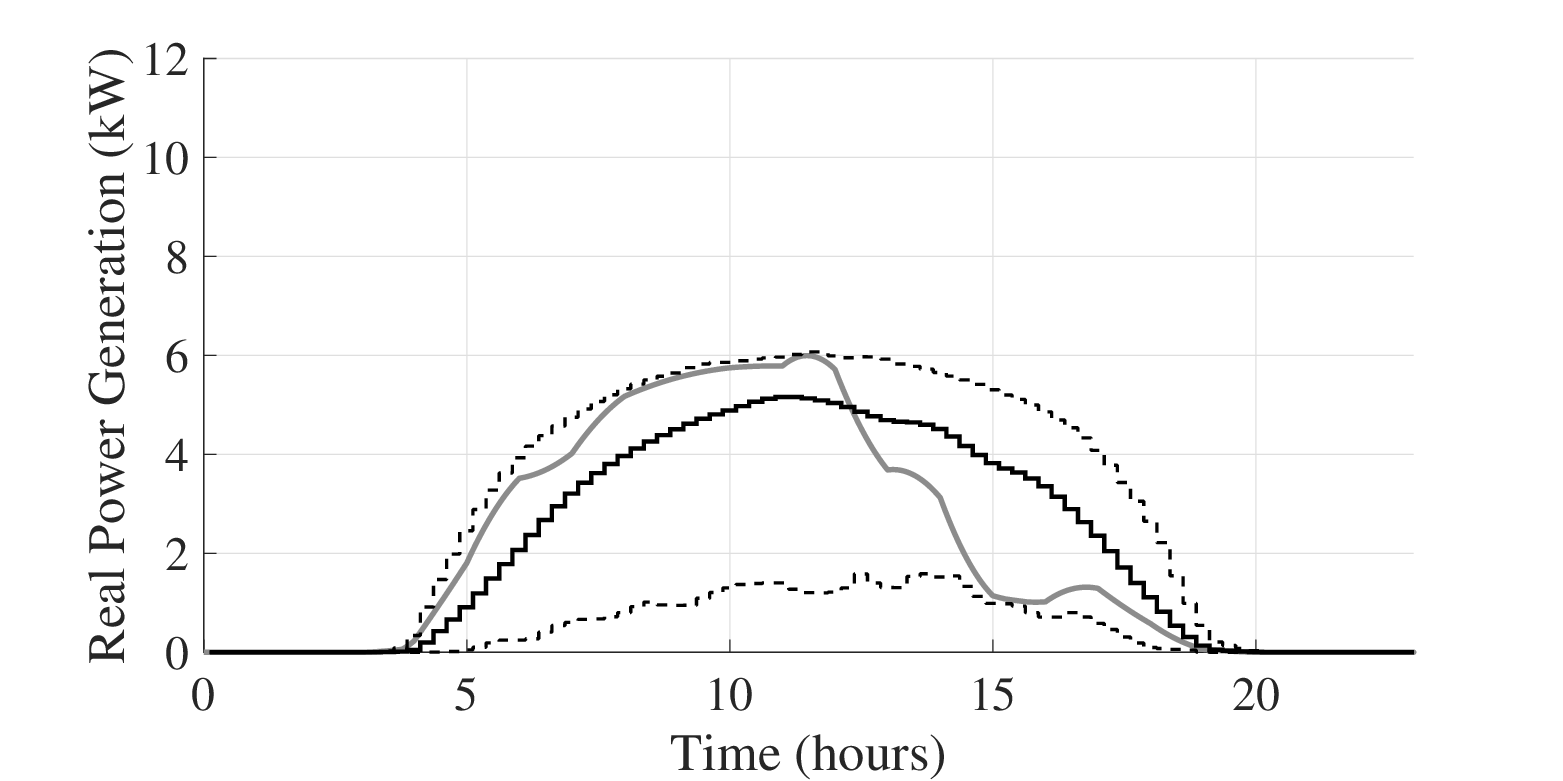}};
    \node[inner sep=0pt] (n1) at (-9,0)
    {\normalsize (a)};
    \node[] (n2) at (-9,-4) {\normalsize (b)};
    \end{tikzpicture}
    \caption{Comparison of actual real power consumption and solar generation for an individual household over a 24-hour period with 15-minute forecasted average, minimum, and maximum.}
    \label{fig:load_solar_forecast}
\end{figure}

Fig.~\ref{fig:load_solar_forecast} compares the 15-minute forecasts generated from the synthetic historical data to the actual load and solar generation profiles for the day of interest for this study.\footnote{Note that the actual solar generation profile shown in Fig.~\ref{fig:load_solar_forecast} is based on solar irradiance data for a typical meterological year (available with GridLAB-D). Thus, this simulation does not include the large fluctuations that are typically seen in real solar irradiance data due to cloud cover.}
For this household, the actual load and solar profiles deviate significantly from the average forecasts, while the worst-case minimum and maximum forecasts reasonably capture the uncertainty caused by load switching and variable solar generation.
Note that if a limited amount of historical data is available, there may be low-probability cases that are not captured by these worst-case forecasts. 
This is illustrated by the 10~kW spike around $t=17$ hours in Fig.~\ref{fig:load_solar_forecast}, which exceeds the maximum load forecast.
This spike occurs due to the air conditioner and water heater switching on simultaneously.

\subsection{Case Study Results}
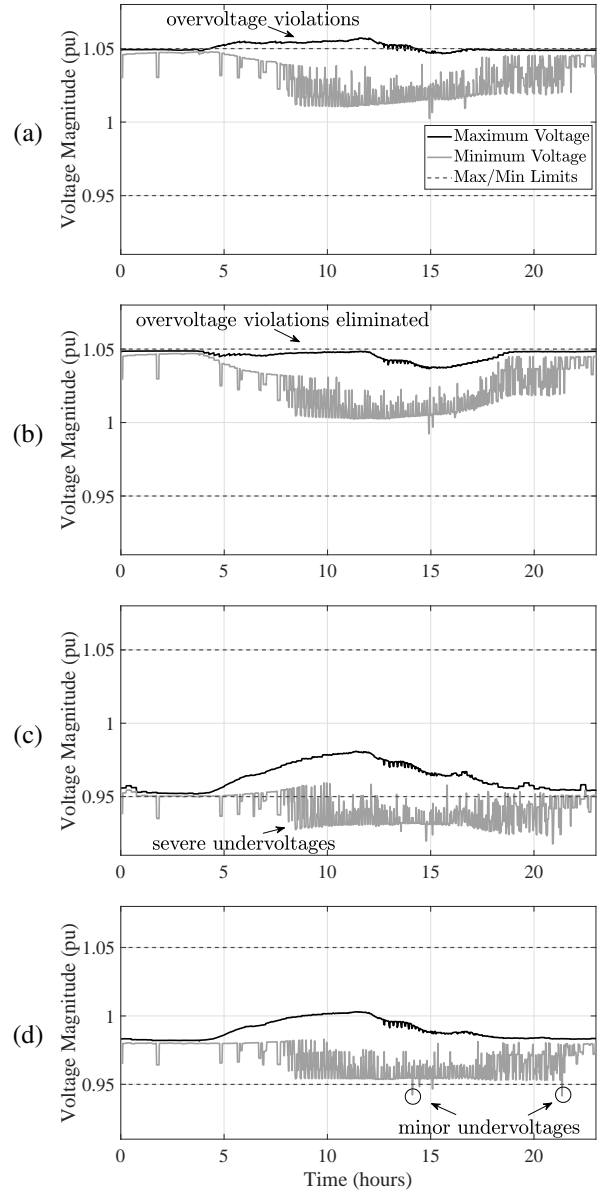
\begin{figure}
    \begin{tikzpicture}
    \node[inner sep=0pt] (n1) at (-5,0)
    {\includegraphics[width=0.45\textwidth,trim=0 0 0 0, clip=true]{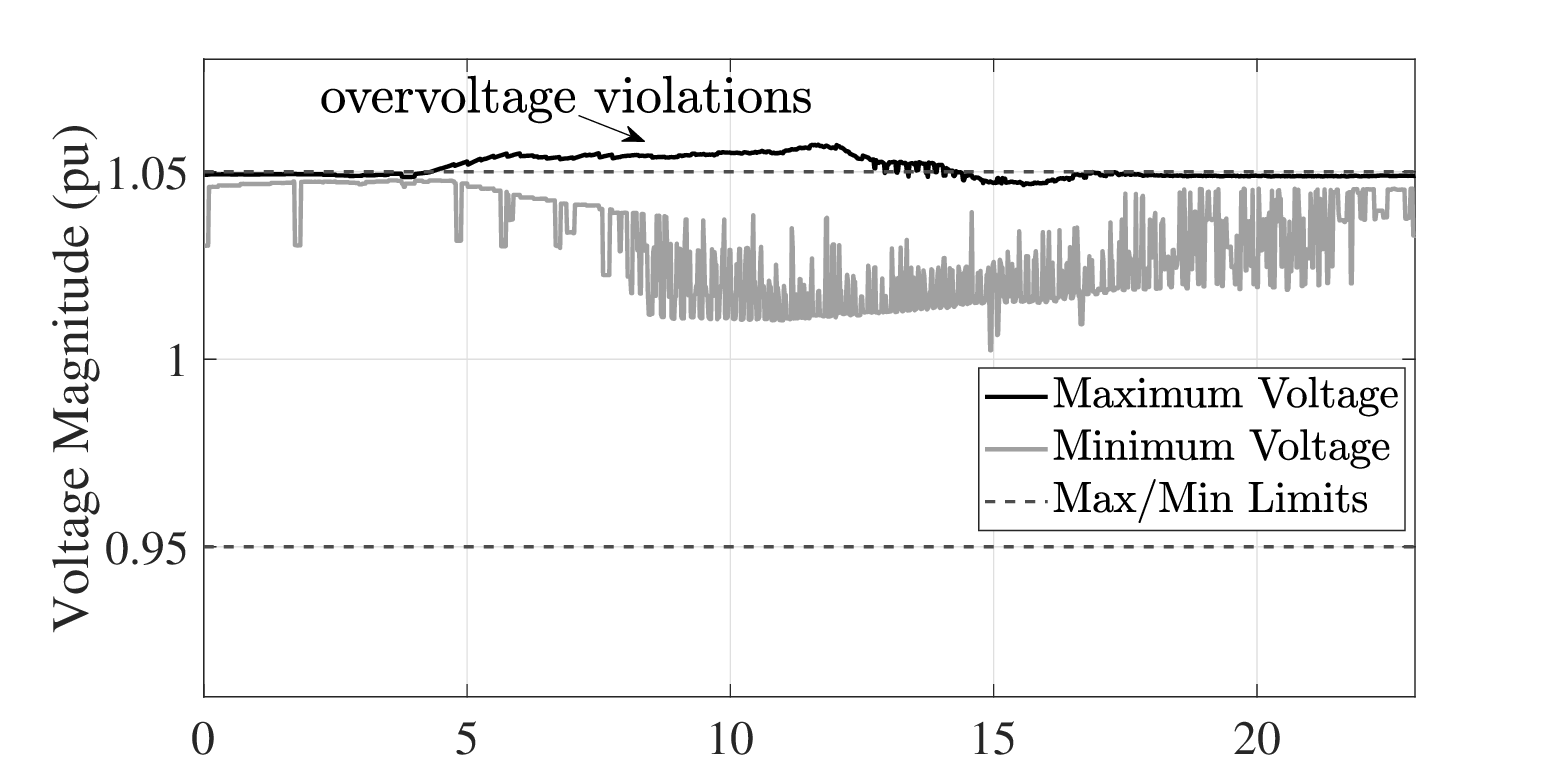}};
    \node[inner sep=0pt] (n1) at (-5,-4)
    {\includegraphics[width=0.45\textwidth,trim=0 0 0 0, clip=true]{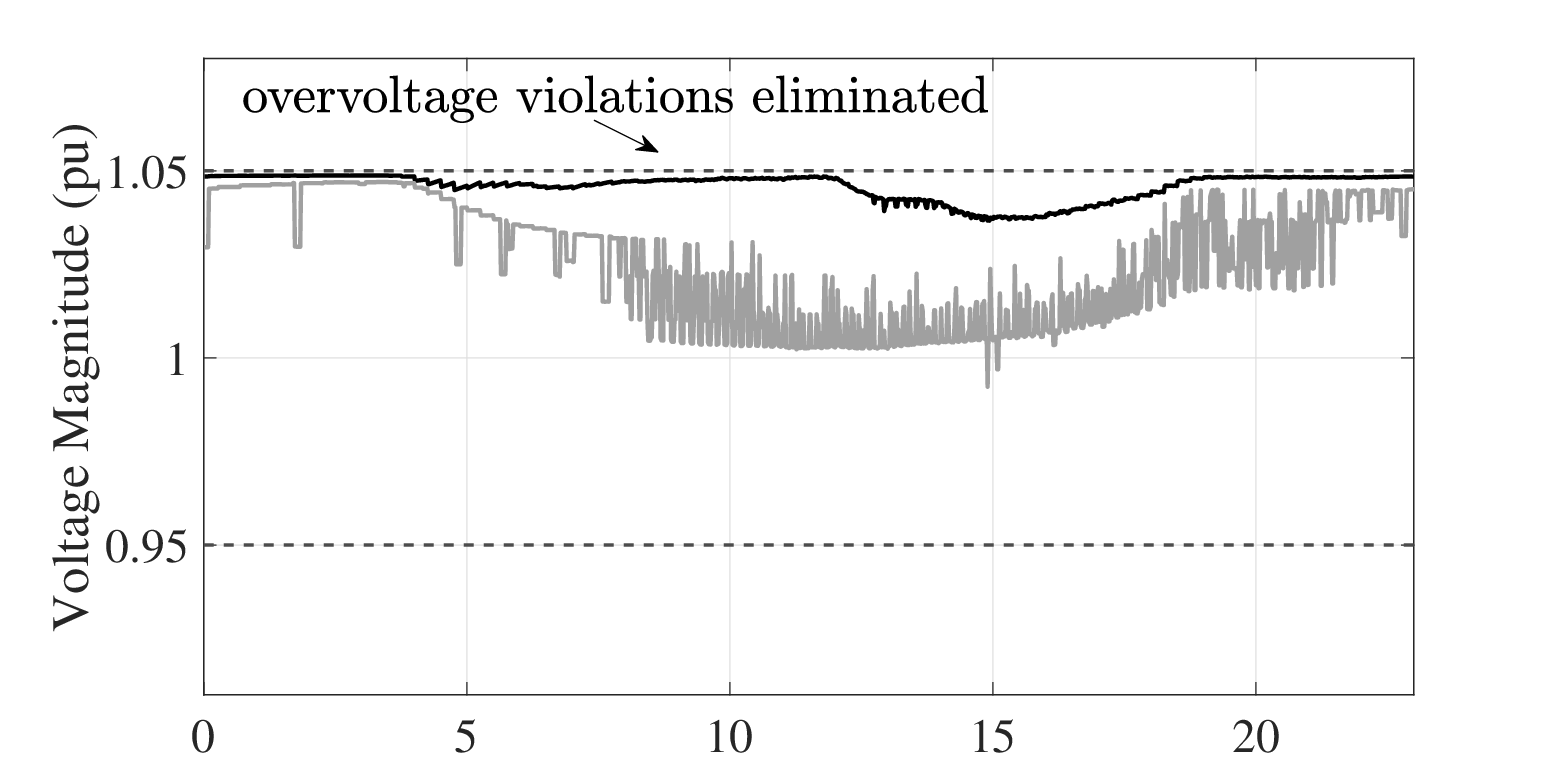}};
    \node[inner sep=0pt] (n1) at (-5,-8)
    {\includegraphics[width=0.45\textwidth,trim=0 0 0 0, clip=true]{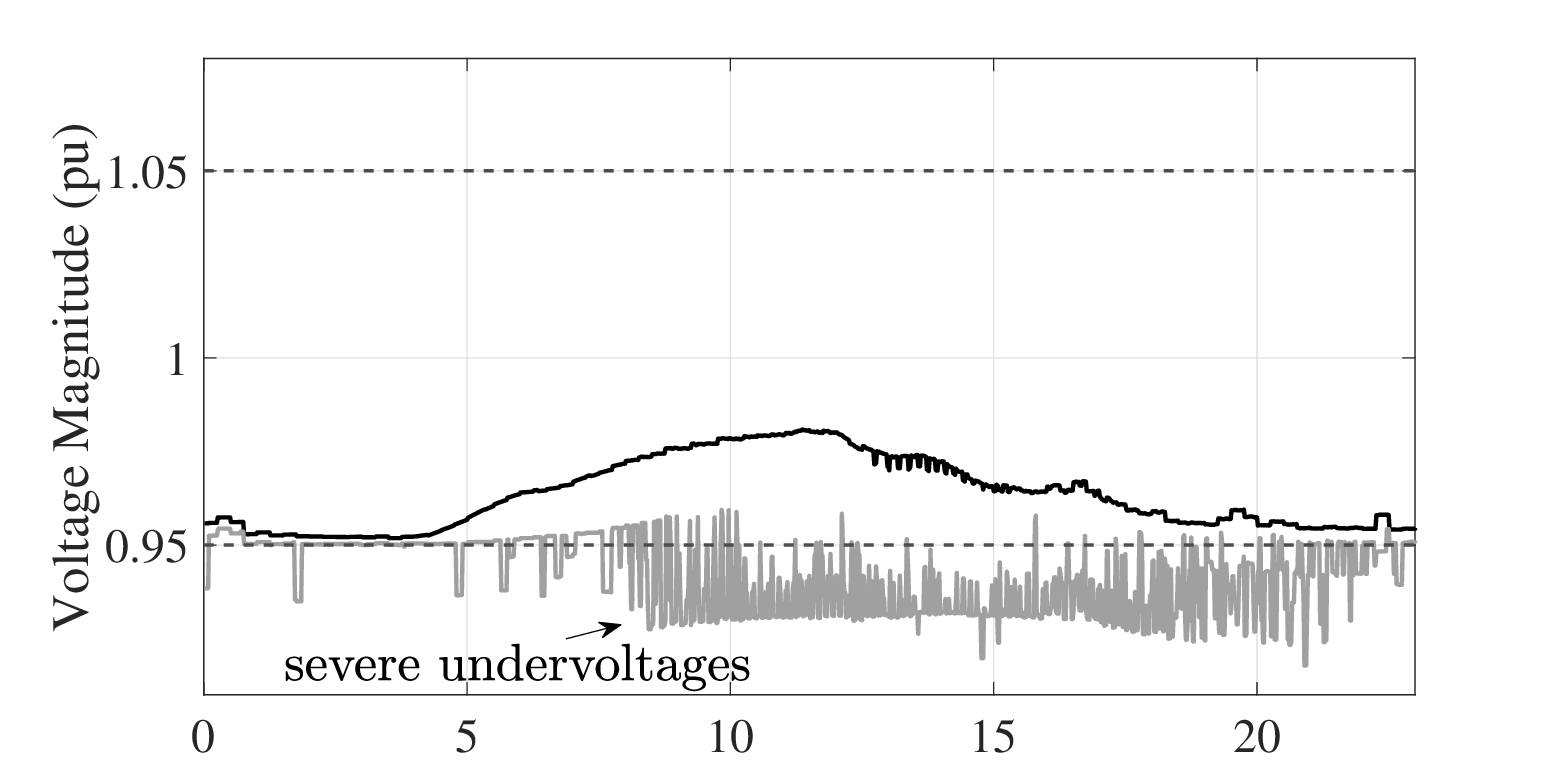}};
    \node[inner sep=0pt] (n1) at (-5,-12)
    {\includegraphics[width=0.45\textwidth,trim=0 0 0 0, clip=true]{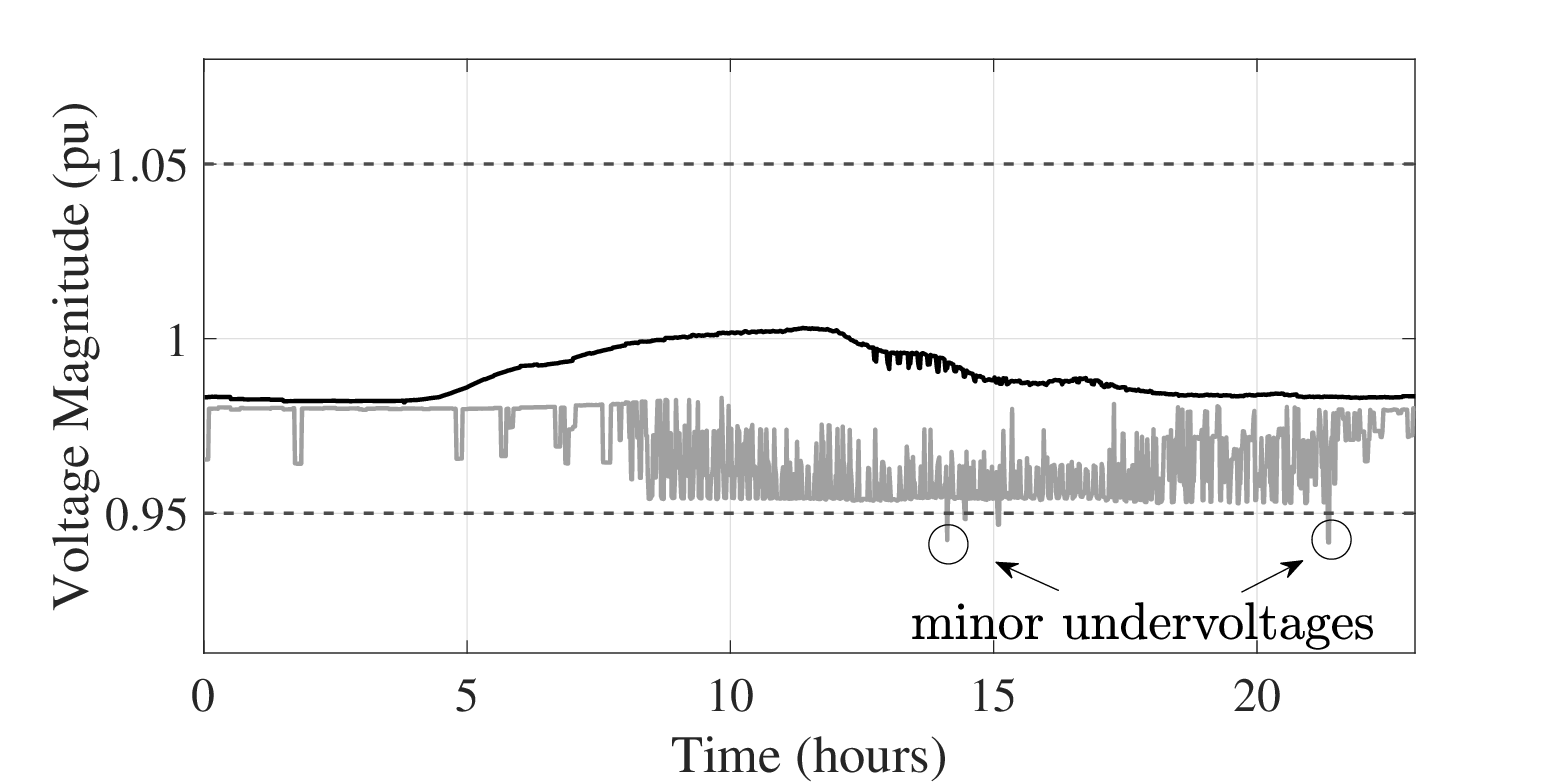}};
    \node[] (n2) at (-9.25,0) {\normalsize (a)};
    \node[] (n2) at (-9.25,-4) {\normalsize (b)};
    \node[] (n2) at (-9.25,-8) {\normalsize (c)};
    \node[] (n2) at (-9.25,-12) {\normalsize (d)};
    \end{tikzpicture}
    \caption{Maximum and minimum voltage magnitudes in the feeder over a 24-hour window. (a) Default voltage constraints and no transformer core losses. (b) Dynamic voltage constraints and no transformer core losses. (c) Default voltage constraints and transformer core losses included. (d) Dynamic voltage constraints and transformer core losses included.}
    \label{Fig:main_results}
\end{figure}

To evaluate the performance of the proposed method, we consider four different variations of the OPF formulation, with and without accounting for forecasting uncertainty and the inclusion of models of transformer core losses. 
Fig.~\ref{Fig:main_results} shows the envelope of voltage magnitudes (i.e., the maximum and minimum voltage magnitudes in each 15-minute window) in the feeder over 24 hours for each of these scenarios.

\begin{figure}
    \begin{tikzpicture}
    \node[inner sep=0pt] (n1) at (-5,0)    {\includegraphics[width=0.45\textwidth,trim=0 0 0 0, clip=true]{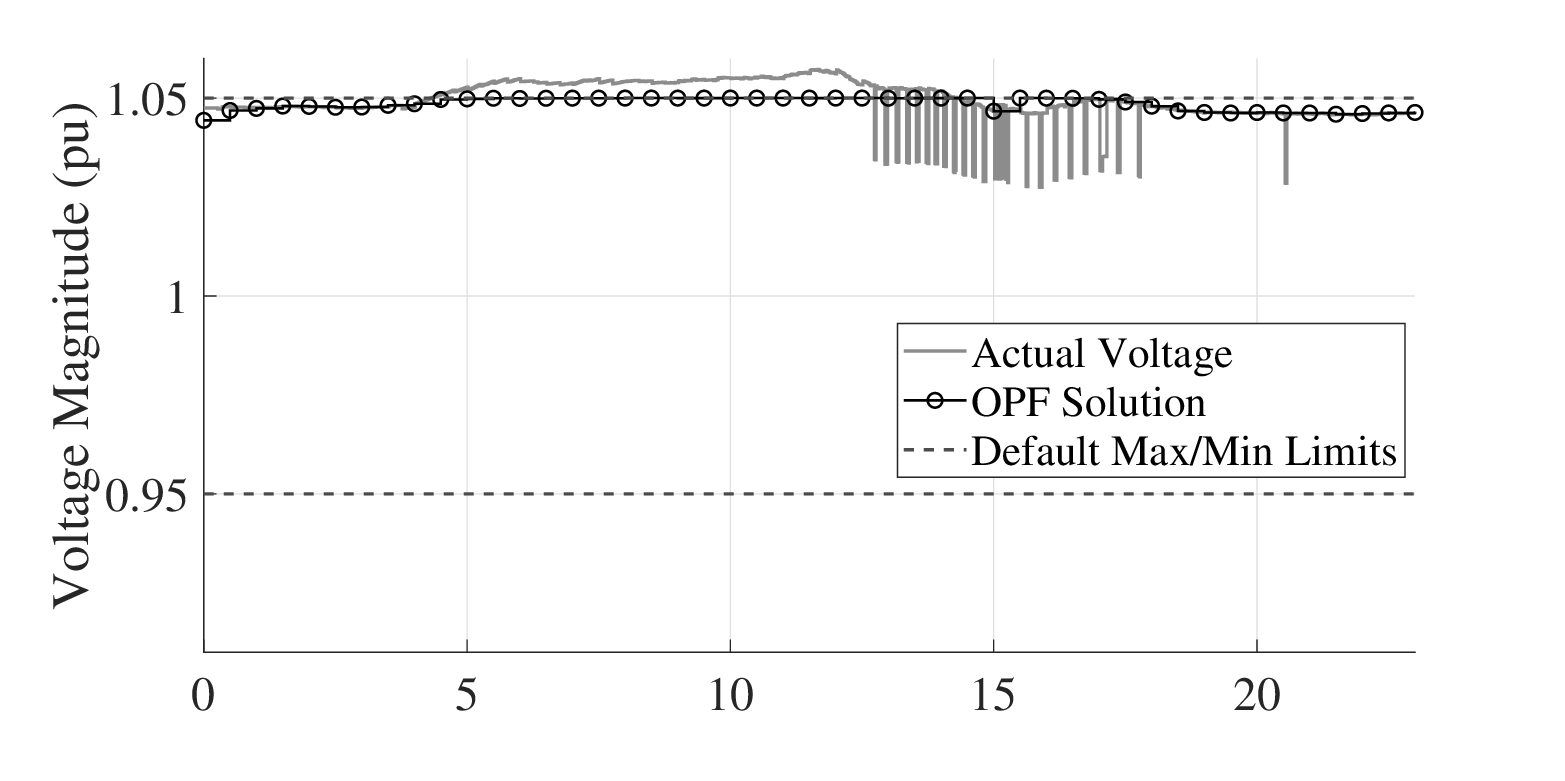}};
    \node[inner sep=0pt] (n1) at (-5,-4)    {\includegraphics[width=0.45\textwidth,trim=0 0 0 0, clip=true]{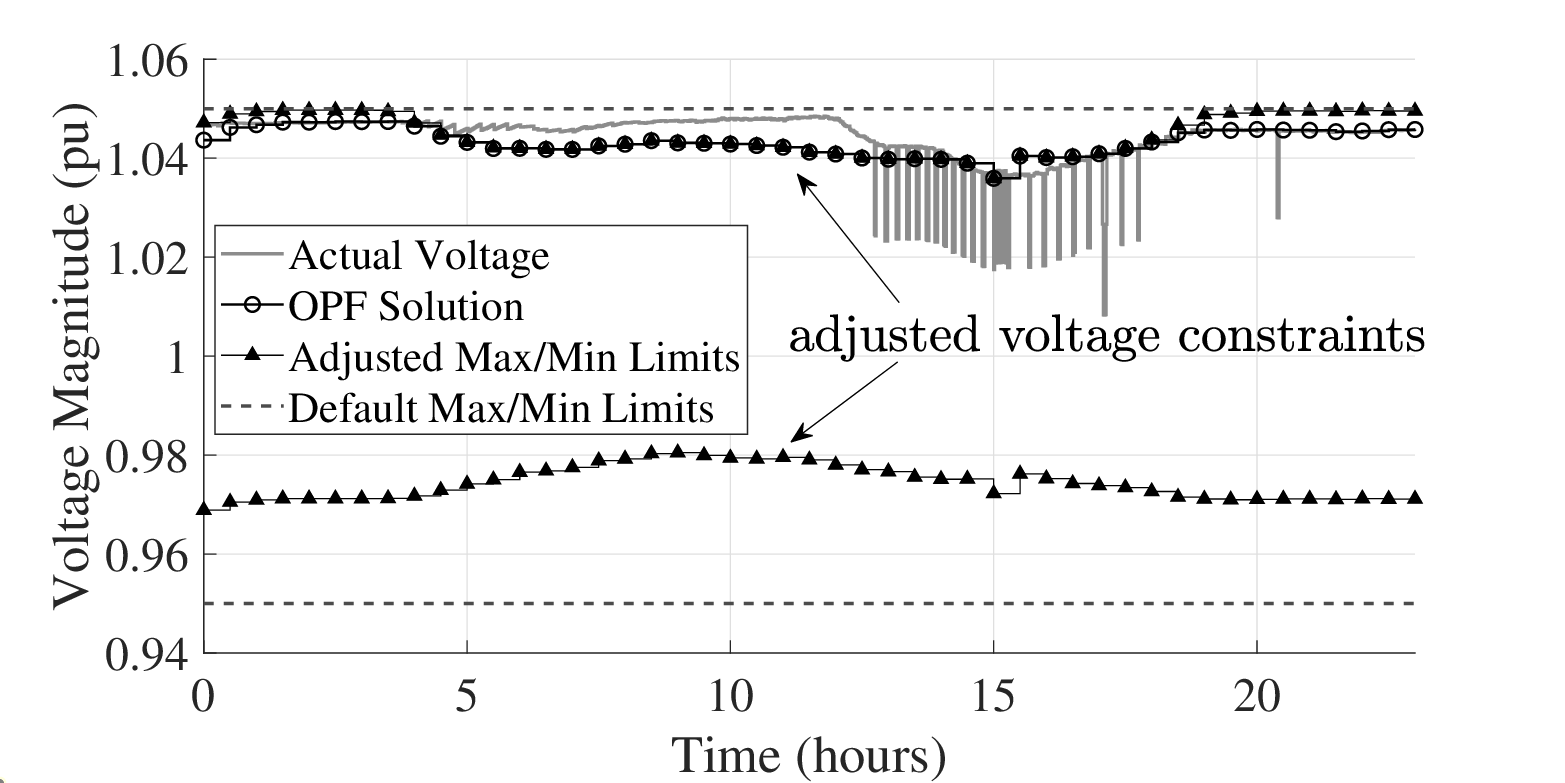}};
    \node[inner sep=0pt] (n1) at (-9.25,0)
    {\normalsize (a)};
    \node[] (n2) at (-9.25,-4) {\normalsize (b)};
    \end{tikzpicture}
    \caption{Comparison of actual voltage magnitudes with OPF solution neglecting transformer core losses for an individual house. (a) OPF with default voltage constraints. (b) OPF with dynamic voltage constraints.}
    \label{Fig:ind_volt_a_b}
\end{figure}

In Figs.~\ref{Fig:main_results}(a) and \ref{Fig:main_results}(b), the core losses of the split-phase center-tapped transformers are neglected in the OPF formulation.\footnote{Note that while core losses are not included in the OPF formulation in Figs.~\ref{Fig:main_results}(a) and~\ref{Fig:main_results}(b), they are included in the GridLAB-D simulation model.}
In order to minimize losses on the distribution lines, the OPF solution tends to push voltage magnitudes towards their upper limit.
In Fig.~\ref{Fig:main_results}(a), the voltage constraints in the OPF are left at their default values of 0.95 and 1.05~pu~\cite{ANSI_C84_1}, and the average load and DER generation forecasts are used.
During the window from $t = 5$~hours to $t = 12$~hours, the actual solar generation exceeds the  forecast  for some houses; see Fig.~\ref{fig:load_solar_forecast}(b).
This causes the actual voltage magnitudes at these nodes to be higher than those predicted by the OPF solution, and results in overvoltage violations, as shown in Fig.~\ref{Fig:ind_volt_a_b}(a).
In contrast, Fig.~\ref{Fig:main_results}(b) shows the system voltage magnitudes when the uncertainty in forecasts due to load switching behavior and variable solar generation are accounted for, as described in Sec.~\ref{sec:jacobian}\@.
By dynamically adjusting the voltage constraints in the OPF based on uncertainty in the forecasts, the overvoltage violations due to deviations in solar generation are eliminated.
Fig.~\ref{Fig:ind_volt_a_b}(b) demonstrates how the voltage constraints in the OPF become more conservative when the load and solar generation forecast uncertainty increases. 
For example, the maximum voltage constraint at  $t = 10$~hours is adjusted to 1.04~pu,  whereas  the actual voltage equals 1.049~pu.
Therefore, even though the actual voltage exceeds the OPF constraint when the solar generation is higher than forecast, the  voltage limit of 1.05~pu is not violated.

\begin{figure}
    \begin{tikzpicture}
    \node[inner sep=0pt] (n1) at (-5,0)    {\includegraphics[width=0.45\textwidth,trim=0 0 0 0, clip=true]{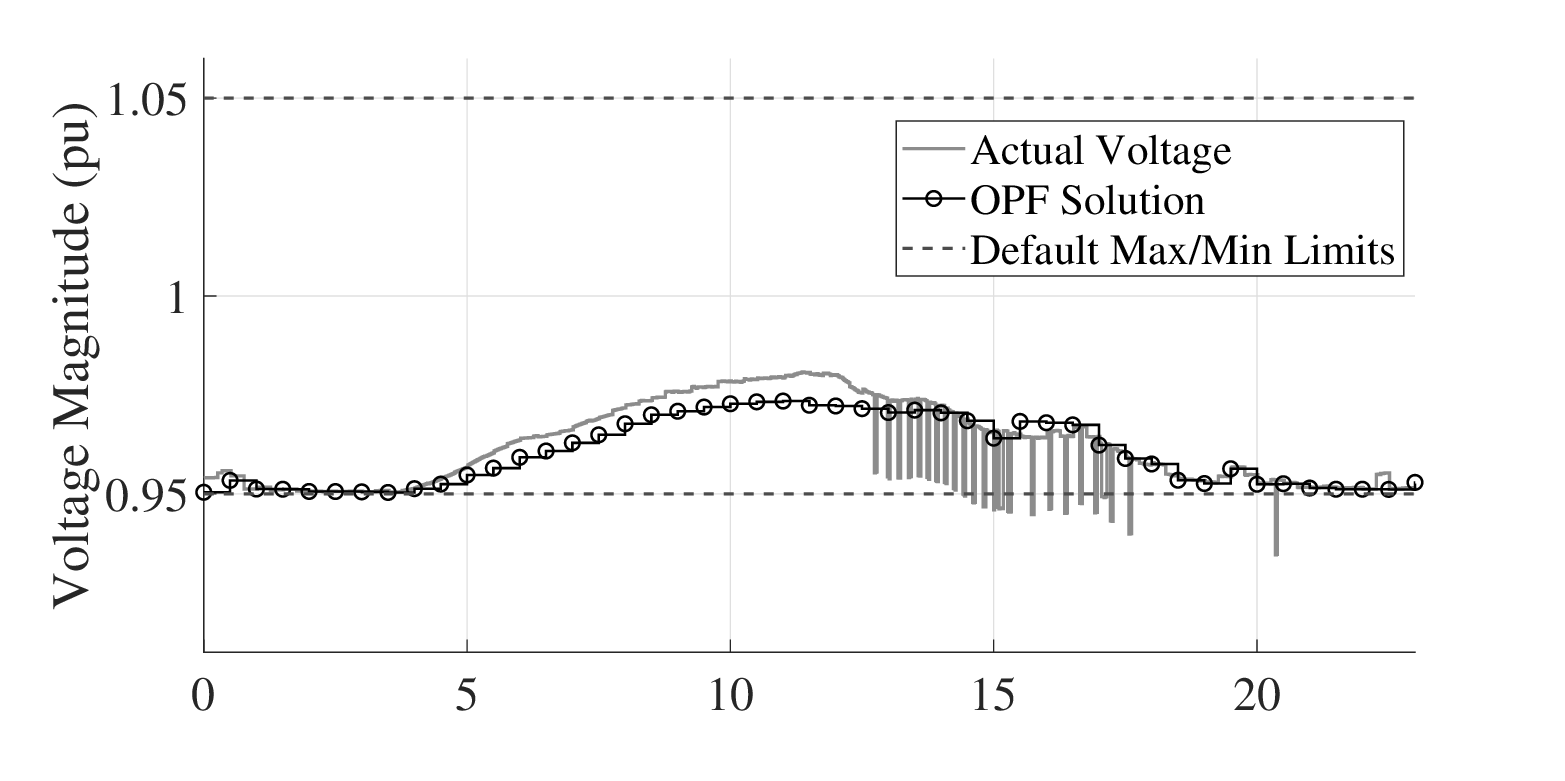}};
    \node[inner sep=0pt] (n1) at (-5,-4)    {\includegraphics[width=0.45\textwidth,trim=0 0 0 0, clip=true]{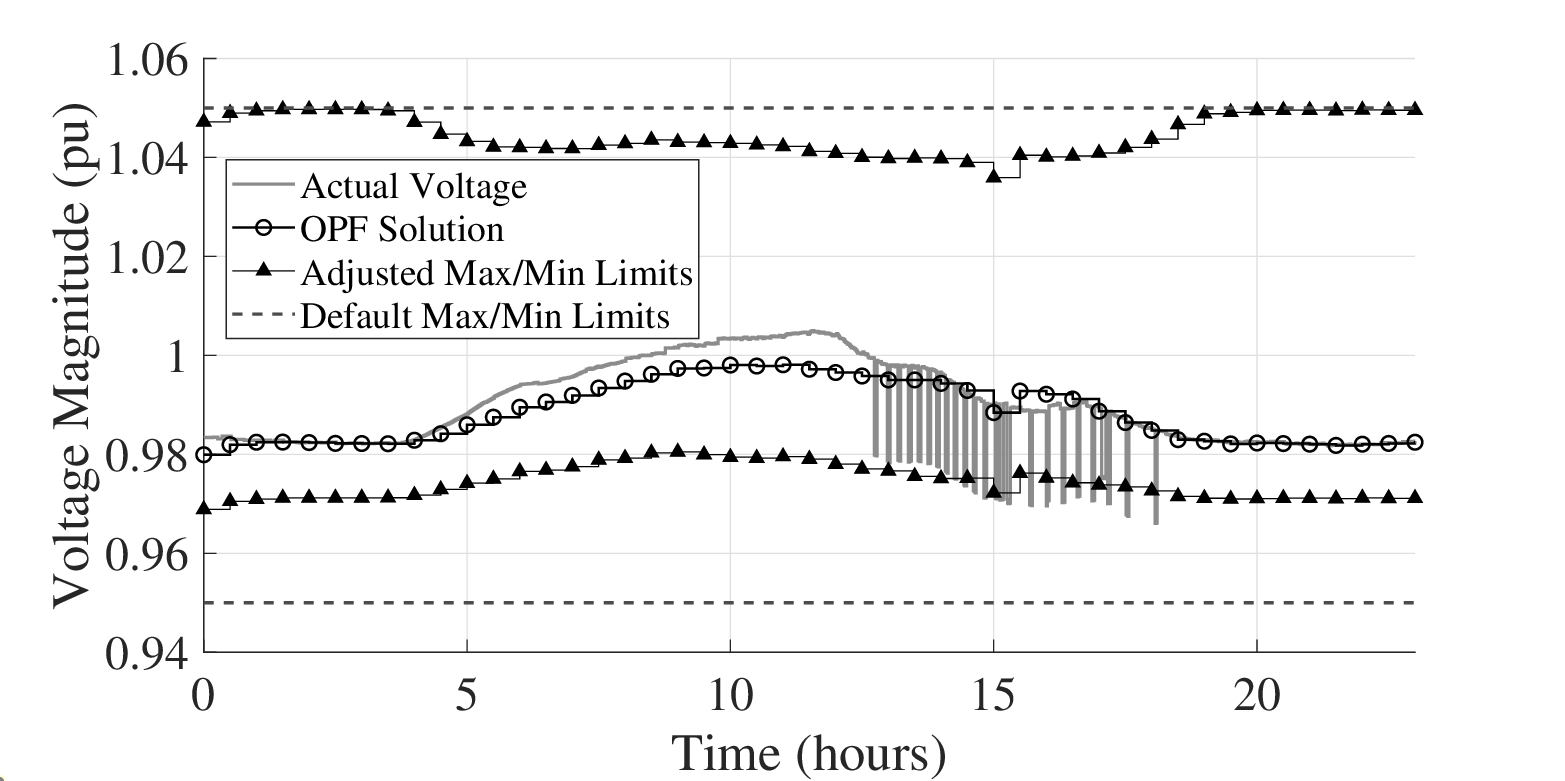}};
    \node[inner sep=0pt] (n1) at (-9.25,0)
    {\normalsize (a)};
    \node[] (n2) at (-9.25,-4) {\normalsize (b)};
    \end{tikzpicture}
    \caption{Comparison of actual voltage magnitudes with OPF solution including transformer core losses for an individual house. (a) OPF with default voltage constraints. (b) OPF with dynamic voltage constraints.}
    \label{Fig:ind_volt_c_d}
\end{figure}

In Figs.~\ref{Fig:main_results}(c) and \ref{Fig:main_results}(d), the core losses of the split-phase center-tapped transformers are included in the OPF formulation by modeling them as constant impedance loads; see Sec.~\ref{sec:split_phase}.
In contrast to the losses on the distribution cables (which decrease at higher voltages), the transformer core losses increase with voltage magnitude.
It turns out that, in this system, the transformer loss dominates that of the lines, hence the OPF solution tends to push voltage magnitudes toward their lower limit when transformer core losses are modeled.
In Fig.~\ref{Fig:main_results}(c), the default voltage limits are used in the OPF\@.
This results in severe undervoltage violations as thermostatically-controlled loads switch on, and the actual power consumption of the houses vary from the OPF voltage solutions based on the average forecast; see Fig.~\ref{Fig:ind_volt_c_d}(a).
However, in Fig.~\ref{Fig:main_results}(d), the uncertainty caused by this switching behavior is accounted for in the OPF by dynamically adjusting the voltage constraints, as shown in Fig.~\ref{Fig:ind_volt_c_d}(b).
This causes the OPF solution to slightly raise the voltage profile of the feeder, and the number and severity of undervoltage violations are drastically reduced.
Note that there are a few instances in Fig.~\ref{Fig:main_results}(d) where the lower voltage limit is still violated even with the proposed approach.
This is due to the worst-case forecasts not capturing low-probability events, as discussed in Sec.~\ref{sec:forecasts}.
However, these voltage violations are small and only last for about a minute, so they may be acceptable in practice.
There are several ways to mitigate these violations if desired, including increasing the value of $\kappa$, using more historical data, or leveraging more advanced statistical techniques for forecasting.

Table~\ref{tab:metric} summarizes the performance of each of the four OPF formulations with respect to the severity and frequency of voltage violations.
Also listed are the net energy supplied by the substation, corresponding to the objective function~\eqref{eq:cost_fun}, and the total system losses (in distribution cables and transformer cores)   over the 24-hour window.
Note that when the default voltage constraints are used in the OPF and forecasting uncertainty is not accounted for, voltages violate their limits for several hours of the day.
Conversely, when voltage constraints are dynamically adjusted based on worst-case forecasts, the severity and duration of voltage violations are significantly reduced.
Finally, by directly modeling the transformer core loss, the proposed approach is able to reduce the net energy supplied by the substation by 6.6\% while maintaining voltages within their limits.
\begin{table}[ht]
\centering
\caption{Performance Metrics for Various OPF Formulations}
\renewcommand{\arraystretch}{1.125}
\begin{tabular}{ |p{2.75cm}||c|c|c|c|}
 \hline
 Includes Core Losses& No & No & Yes & Yes \\
 \hline
 Voltage Constraints & Default & Dynamic & Default & Dynamic \\
 \hline
 Frequency of constraint violations (min)  & 591   & 0 & 866 &  7 \\
 \hline
Severity of constraint violations (pu)  &  +0.007  & 0 & $-$0.032 & $-$0.009 \\
 \hline
Net energy supplied by substation (MWh)  &  4.42 & 4.39 &   3.97 &   4.10  \\
\hline
Total losses (kWh) &  214.25 & 212.91 &  187.90  &  195.30 \\
\hline
\end{tabular}
\label{tab:metric}
\end{table}

\section{Conclusions}
\label{sec:conclusion}

In this work, we addressed two of the unique challenges with implementing the OPF for a distribution system modeled down to the point of connection of individual buildings.
We illustrated how the robustness of the OPF to the switching behavior of residential loads and variable renewable generation can be improved through the dynamic adjustment of voltage constraints.
We also proposed a methodology for including detailed models of split-phase secondaries and center-tapped transformer core losses in a computationally tractable OPF formulation.
The performance of the proposed approach was validated through numerical simulations of a large-scale, realistic distribution feeder.

One potential avenue for future research could be the extension of the proposed approach to other OPF constraints (e.g., current or power constraints on lines and transformers).
Extension of the proposed approach to include other distribution system components such as tap-changing transformers, voltage regulators, or inverters that operate in control modes other than constant PQ (e.g., constant power factor, Volt-VAR) would also be of interest.
Additionally, an evaluation of the scalability of the proposed approach for larger distribution networks would be valuable.

Finally, as we have discussed, the performance of the proposed approach relies on the accuracy of worst-case load and solar generation forecasts based on historical AMI data.
Thus, future research could consider an investigation of the best statistical methods for generating these forecasts.

\bibliographystyle{IEEEtran}
\bibliography{IEEEabrv,references}
%




\end{document}